\documentclass[pra,superscriptaddress,nofootinbib,twocolumn,notitlepage,showpacs]{revtex4-1}

\usepackage{amsmath,graphicx,epsfig,color,amsfonts,amssymb,bbm}
\usepackage{amsthm}
\usepackage{mathtools,color}

\usepackage[caption=false]{subfig}
\let\originaleqref\eqref
\renewcommand{\eqref}{Eq.~\originaleqref}

\newcommand{\ket}[1]{|#1\rangle}

\newcommand{\id}{\mathbbm{1}}
\renewcommand{\S}{\mathcal{S}}

\newcommand{\SexpI}[1]{\mathcal{S}^{\mbox{\tiny exp}}_{#1}}
\newcommand{\Smax}{\mathcal{S}_{\mbox{\tiny max}}}
\newcommand{\Imax}{\mathcal{I}^{\mbox{\tiny max}}}
\newcommand{\ImaxN}[1]{\Imax_{#1}}
\newcommand{\IminN}[1]{\I^{\mbox{\tiny min}}_{#1}}

\newcommand{\vecP}{\vec{P}}
\newcommand{\Pexp}{\vec{P}_{\mbox{\tiny exp}}}

\newcommand{\vcr}{v_{\mbox{\tiny Cr}}}
\newcommand{\nucr}{\nu_{\mbox{\tiny Cr}}}
\newcommand{\NS}{\mathcal{NS}}

\newcommand{\Q}{\mathcal  Q}
\newcommand{\I}{\mathcal  I}
\newcommand{\As}{{\mbox{\tiny A}}}
\newcommand{\Bs}{{\mbox{\tiny B}}}

\newcommand{\proj}[1]{\left| #1 \right\rangle\!\!\left\langle #1 \right|}
\DeclareMathOperator{\tr}{tr}

\usepackage{hyperref}

\renewcommand{\L}{\mathcal{L}}
  
\usepackage[outdir=./]{epstopdf}

\definecolor{nblue}{rgb}{0.2,0.2,0.7}
\definecolor{ngreen}{rgb}{0.2,0.6,0.2}
\definecolor{nred}{rgb}{0.9,0.2,0.2}
\definecolor{nblack}{rgb}{0,0,0}

\newcommand{\red}{\color{nred}}

\begin{document}

\title{Bipartite Bell inequalities with three ternary-outcome measurements \\--- from theory to experiments {\red PLUS A CORRIGENDUM}}

\author{Sacha Schwarz}
\affiliation{Institute of Applied Physics, University of Bern, 3012 Bern, Switzerland.}
\author{B\"anz Bessire}
\affiliation{Institute of Applied Physics, University of Bern, 3012 Bern, Switzerland.}
\author{Andr\'e Stefanov}
\affiliation{Institute of Applied Physics, University of Bern, 3012 Bern, Switzerland.}
\author{Yeong-Cherng~Liang}
\email{ycliang@mail.ncku.edu.tw}
\affiliation{Department of Physics, National Cheng Kung University, Tainan 701, Taiwan.}
\affiliation{Institute for Theoretical Physics, ETH Zurich, 8093 Zurich, Switzerland.}

\date{\today}

\begin{abstract}
We explore quantum nonlocality in one of the simplest bipartite scenarios. Several new facet-defining Bell inequalities for the $\{[3\,3\, 3]\, [3\, 3\, 3]\}$ scenario are obtained with their quantum violations analyzed in details. Surprisingly, all these inequalities involving only genuine ternary-outcome measurements can be violated maximally by some two-qubit entangled states, such as the maximally entangled two-qubit state. This gives further evidence that in analyzing the quantum violation of Bell inequalities, or in the application of the latter to  device-independent quantum information processing tasks, the commonly-held wisdom of equating the local Hilbert space dimension of the optimal state with the number of measurement outcomes is not necessarily justifiable. In addition, when restricted to the {\em minimal} qubit subspace, it can be shown that one of these Bell inequalities requires non-projective measurements to attain maximal quantum violation, thereby giving the first example of a facet-defining Bell inequality where a genuine positive-operator-valued measure is relevant.  We experimentally demonstrate the quantum violation of this and two other Bell inequalities for this scenario using energy-time entangled photon pairs. Using the obtained measurement statistics, we demonstrate how characterization of the underlying resource in the spirit of device-independence, but supplemented with auxiliary assumptions, can be achieved. In particular, we discuss how one may get around the fact that, due to finite-size effects, raw measurement statistics typically violate the non-signaling condition.
\end{abstract}

\maketitle

\section{Introduction}
In the classic paper where Schr\"odinger~\cite{Schroedinger:1935} introduced the term quantum entanglement, he remarked that this is not {\em one} but rather {\em the} characteristic trait of quantum mechanics that forces its entire departure from a classical line of thought. Indeed, among the many nonclassical features offered by entanglement, {\em quantum nonlocality} --- the fact that (certain) entangled quantum systems can exhibit correlations between measurement outcomes that are not Bell-local~\cite{Bell:1964,Brunner:RMP} --- has not only called for a closer inspection of notions like realism, determinism etc., but has also led to the reexamination of the causal structure underlying our physical world~\cite{Bell:Book}.

While the peculiarity of quantum nonlocality has made it more challenging for us to gain good intuitions in the quantum world, the very same feature has also led to quantum information tasks that cannot be achieved otherwise. A prominent example of this is the possibility to perform quantum key distributions whose security is guaranteed without relying on any assumption about the measurements being performed nor the quantum state prepared~\cite{DIQKD,DIQKD2}. Similarly, quantum nonlocality is also an essential ingredient for the self-testing~\cite{mayers,McKague:JPA:2012,Yang:PRA:2013,Swap,SelfTesting} of quantum apparatus directly from measurement statistics. More recently, the paradigm of {\em device-independent quantum information}~\cite{Brunner:RMP,Scarani:DIQIP} --- where the analysis of quantum information is based solely on the observed correlations --- has also been applied in the context of randomness expansion~\cite{BIV:Randomness,rand_colbeck}, randomness extraction~\cite{Chung:arXiv:1402.4797}, dimension-witnessing~\cite{DimWitness,Navascues:PRX,Navascues:PRA:2015}, as well as robust certification~\cite{DIEW,Pal:DIEW,GUBI}, classification~\cite{Brunner:PRL:110501} and quantification~\cite{Moroder:PRL:PPT,Toth:1409.3806,DIWED} of (multipartite) entanglement etc.

For all these tasks, an imperative step is to certify that the observed correlation is not Bell-local --- a task that is often achieved through the violation of {\em Bell inequalities}~\cite{Bell:1964}. Achieving a solid understanding of the quantum violation of Bell inequalities is thus an important step towards the development of novel device-independent quantum information processing tasks. To date, however, the bulk of such studies have focussed on the simplest Clauser-Horne-Shimony-Holt~\cite{CHSH} (CHSH) Bell scenario, namely, one involving only two parties, each performing two binary-outcome measurements. While more complicated Bell scenarios, such as those involving more parties~\cite{WWZB,DIWED,Sliwa:PLA:2003}, or more measurement settings~\cite{Chained,Avis:JPA:2005,Avis:JPA:2006,Brunner:PLA:2008,Pal:PRA:022120} or more measurement outcomes~\cite{CGLMP,Kaszlikowski:PRA:2002,Collins:JPA:2004} have also been individually considered, scenarios involving a combination of these have so far received relatively little attention (see, however, Refs.~\cite{Massar:PRA:052112,BKP,GUBI,Grandjean:PRA:2012,Collins:JPA:2004}).

In this paper, we investigate the Bell scenario of two parties where each experimenter can perform three ternary-outcome measurements. 
Although some Bell inequalities in this scenario have previously been reported~\cite{Buhrman:PRA:052103,Massar:PRA:052112,BKP,Collins:JPA:2004}, most of these are not facet-defining~\cite{Avis:JPA:2005} for the corresponding convex set of Bell-local correlations. In contrast, we present in this work several novel facet-defining Bell inequalities for this scenario. Interestingly, some of these newly obtained Bell inequalities --- despite being ternary-outcome and irreducible to one having fewer measurement outcomes --- can already be violated maximally via local measurements on  entangled two-qubit states. Our work thus complements that of Refs.~\cite{Pal:042105,Pal:PRA:022120,Grandjean:PRA:2012},  showing that in determining the  quantum state that maximally violates a given Bell inequality, optimal choice of the local Hilbert space dimension is not necessarily correlated with the (maximal) number of measurement outcomes involved.

A common feature shared by all the Bell inequalities that we present here is that they {\em cannot} be cast in a form involving only full (bipartite) correlators~\cite{GUBI,Bancal:Symmetric,MGS}. Interestingly, except in Bell experiments~\cite{BellExp,LoopholeFreeBellExp} related to the closing of detection loophole~\cite{Larsson:Loopholes}, there is almost no other experimental exploration of this generic kind of Bell inequalities (see however Ref.~\cite{Christensen:arXiv:1506.01649}). In addition, among all these novel inequalities, there is one which provably requires non-projective measurements in order to attain its maximal quantum violation when one restricts to the minimal, qubit subspace. Here, we experimentally violate this and two other Bell inequalities --- all involving three ternary-outcome measurements --- using energy-time entangled photons. Due to strong resistance against decoherence and the possibility to manifest entanglement in different degrees of freedom (polarization \cite{kwiat1999}, transverse or orbital angular momentum \cite{mair2001_tomo,Dada:NatPhys}, or energy-time \cite{law2000,thew2004,richart2012,bernhard2013,bessire2014}), entangled photon pairs offer an ideal framework for such fundamental studies. Energy-time entangled photons, in particular, represent a highly-flexible system tunable using techniques developed from ultrafast science \cite{peer2005,zaeh2008}, especially for the preparation of quantum states with varying degree of entanglement and/or Hilbert space dimensions.

The rest of this paper is structured as follows. Notations and other preliminary materials are introduced in section~\ref{Sec:Prelim}. We then present in section~\ref{Sec:BI} the Bell inequalities that we have obtained through numerical optimizations. Experimental Bell inequality violations are reported in section~\ref{sec:ExpDemo}. After that, in section~\ref{Sec:DI}, we discuss about analysis of the measured data along the spirit of device-independent quantum information, using entanglement quantification via negativity~\cite{Vidal:Negativity} as an example. We end with some further discussions in section~\ref{Sec:Conclusion}. Technical details related to numerical optimizations and certain results obtained thereof are relegated to the appendices.

\section{Preliminaries}
\label{Sec:Prelim}

\subsection{Bell inequalities and some natural sets of correlations}

Consider a Bell-type experiment involving two parties Alice and Bob, where each party is allowed to perform three ternary-outcome measurements. We label Alice's measurement setting (input) by $x$, Bob's by $y$ and their corresponding measurement outcome (output) by $a$ and $b$ respectively. For ease of discussion, we follow the notation of Ref.~\cite{Barnea:PRA:2013} and  refer to this as the $\{[3\, 3\, 3]\, [3\, 3\, 3]\}$ Bell scenario, where the number of entries in the first (second) square bracket is the number of input for Alice (Bob) while the actual value in the square brackets represents the number of output for that particular input.

The correlations between measurement outcomes observed in a Bell-type experiment in this scenario can be succinctly summarized using the vector ${\vec{P}}=\{P(a,b|x,y)\}_{x,y,a,b=0}^2$ of joint conditional probabilities. A correlation is said to be Bell-local~\cite{Brunner:RMP,Bell:1964} if it admits the decomposition
\begin{equation}\label{Eq:BellLocal}
	P(a,b|x,y)\stackrel{\L}{=} \sum_{\lambda} P_\lambda P(a|x,\lambda) P(b|y,\lambda)
\end{equation}
for all $x,y,a,b$ with some fixed, normalized weights $P_\lambda\in[0,1]$, where we denote throughout by $\L$ the set of Bell-local correlations. It turns out~\cite{Pitowsky:Book} that $\L$ is a convex polytope~\cite{Polytopes}, and thus can be described by a convex mixture of a finite number of (deterministic) extremal probability vectors satisfying $P(a|x,\lambda)=0,1$ and $P(b|y,\lambda)=0,1$. Equivalently, a convex polytope can be fully characterized by the intersection of a finite number of half-spaces~\cite{Polytopes}. Following Ref.~\cite{Avis:JPA:2005}, we refer to the minimal set of such half-spaces as facet-defining Bell inequalities, or simply as facets. A generic Bell inequality, i.e., one involving some linear combination of joint conditional probabilities, for the $\{[3\, 3\, 3]\, [3\, 3\, 3]\}$ scenario reads as
\begin{equation}\label{Eq:betaS}
	\sum_{x,y,a,b=0}^2 \alpha^{xy}_{ab} P(a,b|x,y) \stackrel{\L}{\le} \Smax^\L\left(\{\alpha^{xy}_{ab}\}\right),
\end{equation}
and is, however, not necessarily a facet. Here, we write explicitly the dependence of $\Smax^\L$ on $ \{\alpha^{xy}_{ab}\}$ to remind that the upper bound attainable by $\vec{P}\in\L$ is a function of the real coefficients $\{\alpha^{xy}_{ab}\}$. 

As was first shown by Bell, the set of correlation vectors $\vecP$  allowed in quantum theory (denoted by $\Q$) is a strict superset of $\L$. Formally, quantum correlations  take the form of 
\begin{subequations}\label{Eq:QuantumProb}
\begin{equation}
	P(a,b|x,y)\stackrel{\Q}{=} \tr(\rho\, M_{a|x}\otimes M_{b|y})
	\label{eq:Pabxy}
\end{equation}
where $\rho$ is a density matrix and $M_{a|x}$ and $M_{b|y}$ are the positive-operator-valued measure (POVM)  elements associated with Alice's and Bob's local measurements, i.e., they satisfy
\begin{equation}
M_{a|x}\ge 0,\,\, M_{b|y}\ge0,\,\, \sum_a M_{a|x}=\id_A,\,\, \sum_b M_{b|y}=\id_B
\end{equation}
\end{subequations}
for all $x,y,a,b$, with $\id_A (\id_B)$ being the identity operator acting on Alice's (Bob's) Hilbert space.

It is easy to see that the marginal distributions of the quantum correlation, $P(a|x,y)$ and $P(b|x,y)$, are independent of the input of the other party, i.e., they satisfy the so-called non-signaling (NS) conditions~\cite{NS,Barrett:PRA:2005}
\begin{equation}\label{Eq:NS}
\begin{split}
	P(a|x,y) &\equiv \sum_{b} P(a,b|x,y) \stackrel{\NS}{=} P(a|x),\\
	P(b|x,y) &\equiv \sum_{a} P(a,b|x,y) \stackrel{\NS}{=} P(b|y).
\end{split}
\end{equation}
Note that if these conditions are violated independent of spatial separation, then Alice can communicate superluminally the value of $x$ to Bob by remotely varying the marginal distribution observed by Bob through her choice of $x$. Interestingly, the set of correlations satisfying \eqref{Eq:NS}, which we shall denote by $\NS$ is actually a strict superset of $\Q$ (see, for instance, Refs.~\cite{Brunner:RMP,AlmostQuantum} and references therein).

Due to the non-signaling nature of $\vecP\in\{\L,\Q\}$, instead of specifying $3^4+1=82$ real coefficients [$\alpha^{xy}_{ab}$ and $\Smax^\L\left(\{\alpha^{xy}_{ab}\}\right)$]  in defining a Bell inequality, one can employ a more compact representation due to Collins and Gisin~\cite{Collins:JPA:2004}, which requires only the specification of $(2\times3+1)^2=49$ parameters. Explicitly, the Collins-Gisin representation of a Bell inequality in this scenario reads as
\begin{equation}\label{Eq:CGForm}
\begin{split}
	\vec{\beta}\cdot\vec{P}=&\sum_{x=0}^2\sum_{a=0}^1 \beta^x_{A,a} P(a|x)+\sum_{y=0}^2\sum_{b=0}^1\beta^y_{B,b} P(b|y)\\ 
	+&\sum_{x,y=0}^2\sum_{a,b=0}^1 \beta^{xy}_{ab} P(a,b|x,y)   \stackrel{\L}{\le}  \Smax^\L({\vec{\beta}}),
\end{split}
\end{equation}
where $\vec{\beta}$ is a vector with entries given by the Bell coefficients $\beta^x_{A,a}$, $\beta^y_{B,b}$, $\beta^{xy}_{ab}$ appearing in \eqref{Eq:CGForm}, while $\vec{P}$ is now the corresponding vector of marginal and joint conditional probability distributions. This particular way of writing a Bell inequality admits the compact tabular representation
\begin{equation}\label{Eq:CGtable}
    \left(
    \begin{array}{r||cc|cc|cc}
    & \beta_{B,0}^0 & \beta_{B,1}^0 & \beta_{B,0}^1 & \beta_{B,1}^1 &\beta_{B,0}^2 & \beta_{B,1}^2 \\ \hline\hline
    \beta_{A,0}^0 &  \beta^{00}_{00} &  \beta^{00}_{01} &  \beta^{01}_{00} &  \beta^{01}_{01} & \beta^{02}_{00} &  \beta^{02}_{01}\\
    \beta_{A,1}^0 &  \beta^{00}_{10} &  \beta^{00}_{11} &  \beta^{01}_{10} &  \beta^{01}_{11} & \beta^{02}_{10} &  \beta^{02}_{11}\\ \hline
    
    \beta_{A,0}^1 &  \beta^{10}_{00} &  \beta^{10}_{01} &  \beta^{11}_{00} &  \beta^{11}_{01} & \beta^{12}_{00} &  \beta^{12}_{01}\\
    \beta_{A,1}^1 &  \beta^{10}_{10} &  \beta^{10}_{11} &  \beta^{11}_{10} &  \beta^{11}_{11} & \beta^{12}_{10} &  \beta^{12}_{11}\\ \hline
    
    \beta_{A,0}^2 &  \beta^{20}_{00} &  \beta^{20}_{01} &  \beta^{21}_{00} &  \beta^{21}_{01} & \beta^{22}_{00} &  \beta^{22}_{01}\\
    \beta_{A,1}^2 &  \beta^{20}_{10} &  \beta^{20}_{11} &  \beta^{21}_{10} &  \beta^{21}_{11} & \beta^{22}_{10} &  \beta^{22}_{11}\\
    \end{array}
    \right)\stackrel{\L}{\le}  \Smax^\L({\vec{\beta}}),
\end{equation}
so that the left-most column are the Bell coefficients associated with Alice's marginal probabilities, the top row gives the Bell coefficients for Bob's marginal probabilities, while each $x$-th block row and $y$-th block column gives the Bell coefficients for the joint distribution of the input combination $(x,y)$.

\subsection{Robustness of quantum violation of Bell inequalities}

With the judicious choice of an entangled quantum state $\rho$ and local measurements described by $M_{a|x}$ and $M_{b|y}$, the resulting quantum correlation $\vec{P}_\Q$ may lead to the violation of a Bell inequality. Given the identities of \eqref{Eq:NS}, it is clear that a Bell inequality can be written in infinitely many different forms. Due to this arbitrariness, the difference between the corresponding quantum value $\S^\Q(\vec{\beta},\rho,\{M_{a|x},M_{b|y}\})$ and the  local bound $\Smax^\L(\vec{\beta})$ is thus not a good figure of merit for comparing different Bell inequalities. Rather, a commonly adopted measure that is unaffected by such an arbitrariness is given by the extent to which the correlation $\vecP_\Q$ can tolerate white noise before it stops violating the given Bell inequality.

Formally, let us denote the uniform probability distribution (white noise) for the Bell scenario $\{[3\, 3\, 3]\, [3\, 3\, 3]\}$ by $\vec{P}_\id$, i.e., ${P}_\id(a,b|x,y)=\frac{1}{3^2}$ for all $a,b,x,y$. Then for any given (nonlocal) quantum correlation $\vec{P}_\Q$ that violates a Bell inequality $\I$, let us consider the convex mixture
\begin{equation}\label{Eq:ConvexCombi}
	\vecP_v=v\vecP_\Q+(1-v)\vec{P}_\id, \quad 0\le v\le 1.
\end{equation}
Clearly, since $\vecP_\Q\not\in\L$, $\vecP_\id\in\L$ and $\L$ is convex,  as we decrease the value of $v$ from 1,
there is some critical value $\vcr^\I$ such that for all $v\in[0,\vcr^\I]$, $\vecP_v$ does not violate $\I$. This critical value $\vcr^\I$ --- which does not depend on how $\I$ is represented --- is commonly referred to as the (white-noise) visibility of $\vecP_\Q$ with respect to $\I$. More generally, as we decrease the value of $v$ from 1, due to the convexity of $\L$, there is some critical value $\vcr$ such that for all $v\in[0,\vcr]$, $\vecP_v\in\L$. Hereafter, we refer to this critical value as the (white-noise) visibility of $\vecP_\Q$ with respect to $\L$. It is worth noting that for any given nonlocal $\vecP_\Q$, $\vcr$ can be efficiently computed using linear programing (see Appendix~\ref{App:LP} for details).

In the event when $\vecP_\Q$ arises from measuring only rank-1 projectors on an entangled two-qutrit state $\rho$, $\vcr^\I$ defined above also coincides with the infimum of $\nu$ in 
\begin{equation}\label{Eq:StateVisibility}
	\rho_\nu=\nu\rho+(1-\nu)\frac{\id}{d^2}, \quad 0\le \nu\le 1,
\end{equation}
before $\rho_\nu$ stops violating $\I$ for the given measurements $\{M_{a|x}, M_{b|y}\}$; in \eqref{Eq:StateVisibility}, $d$ is the local Hilbert space dimension of $\rho$, e.g., $d=3$ in the case of a qutrit. Hereafter, we refer to the infimum of $\nu$ in \eqref{Eq:StateVisibility} for the general scenario, i.e., when $\rho$ is not necessarily a two-qutrit state and/or $\{M_{a|x}, M_{b|y}\}$ are not necessarily rank-1 projectors as the {\em state visibility} $\nucr^\I$ of $\rho$ with respect to $\I$. As it stands, this visibility  depends not only on $\rho$ and $\I$, but also on the choice of POVM elements $\{M_{a|x}, M_{b|y}\}$. The visibilities $\nucr^I$ and $\vcr^I$ will be our main figures of merit in comparing the different Bell inequalities presented in the next section.

\section{New facet-defining inequalities and their quantum violation}
\label{Sec:BI}

\subsection{Searching for new facet-defining Bell inequalities}

Let us now briefly recapitulate the state of the art of various Bell scenarios. For $\{[2\,2]\, [2\, 2]\}$, $\{[3\,3]\, [3\, 3]\}$, $\{[2\,2\, 2]\, [2\, 2\, 2]\}$, $\{[2\,2\, 2]\, [2\, 2\, 2\, 2]\}$, $\{[2\,3]\, [2\, 2\, 2]\}$ and $\{[2\,2]\, [2\, 2]\, [2\, 2]\}$, the complete list of facet-defining Bell inequalities has been obtained with the help of standard polytope softwares, see Refs.~\cite{Sliwa:PLA:2003,Collins:JPA:2004,Pironio:AllCHSH}. Generalizing the results of Refs.~\cite{Sliwa:PLA:2003,Collins:JPA:2004}, Pironio~\cite{Pironio:AllCHSH} recently showed that there is a large family of Bell scenarios where the only non-trivial facets are either the CHSH inequality or their liftings~\cite{Pironio:Lifting}. Beyond this, only partial lists of facet-defining Bell inequalities are known. Specifically, those for $\{[2\,2\, 2\, 2]\, [2\, 2\, 2\, 2]\}$ and  $\{[2\,2\, 2\, 2\, 2]\, [2\, 2\, 2\, 2\, 2]\}$ can be found in Refs.~\cite{Brunner:PLA:2008,Pal:PRA:022120} (the recent work by Deta and Sikiri\'c~\cite{Deta:1501.05407}, however, suggests that the known list of 175 facets~\cite{JD-TV} for $\{[2\,2\, 2\, 2]\, [2\, 2\, 2\, 2]\}$ is complete).

In this paper, we shall restrict our attention to the Bell scenario  $\{[3\,3\, 3]\, [3\, 3\, 3]\}$ where, to our knowledge, the only known (non-lifted) nontrivial facet is the one presented in Ref.~\cite{Collins:JPA:2004}. To get a better idea of what quantum entanglement has to offer in this scenario, we shall first generate some novel facets for this scenario. While a few techniques~\cite{Avis:JPA:2005,Avis:JPA:2006,Pal:PRA:022120}  are known in the literature for generating facets for $\L$, here we adopt a different approach --- based on linear programming --- which allows us to obtain nontrivial facets that can be violated by quantum theory with some nontrivial $\vcr$. It is worth noting that the search for Bell inequality using linear programming has also been considered~\cite{Massar:PRA:052112} in the context of minimizing the detection efficiency requirement in a loophole free Bell test.

Let us now recall from Ref.~\cite{Liang:PRA:2009} the following Bell inequality (first introduced in Ref.~\cite{Buhrman:PRA:052103})
\begin{equation}\label{Ineq:I3plus}
    I_3^+:\S_0 =\frac{1}{9}
    \sum_{x,y,a,b=0}^{2}\delta_{xy+a+b}\,P(a,b|x,y)\stackrel{\L}{\le}\frac{2}{3}
\end{equation}
where $\delta_{f}$ is a short-hand for the Kronecker delta $\delta_{f\,\text{mod}\,3,0}$. This inequality was rediscovered (in a different form) in Ref.~\cite{Ji:PRA:2008} and has  been discussed as a specific case of a family of generalization of the CHSH Bell inequalities~\cite{Liang:PRA:2009,Wang:1109.4988,Bavarian:ICTS,Howard:042103,Ravi:1502.02974,Murta:1510.09210}. Maximal quantum violation ($\approx 0.7124$) of this inequality can be achieved by (locally) performing mutually unbiased  measurements on the maximally entangled two-qutrit state~\cite{Liang:PRA:2009} (see also Ref.~\cite{Ji:PRA:2008}), i.e., 
\begin{equation}
	\ket{\Phi^+_3}=\frac{1}{\sqrt{3}}\sum_{i=0}^2\ket{i}_\As\ket{i}_\Bs,
\end{equation}
where $\{\ket{i}_\As\}$ are orthonormal basis vectors for Alice's (qutrit) Hilbert space; likewise for $\{\ket{i}_\Bs\}$.
This  feature of $I_3^+$ naturally suggests that it may be used for the self-testing of $\ket{\Phi^+_3}$. Indeed, numerical optimization using the tools of Ref.~\cite{Moroder:PRL:PPT} shows that when one approaches the maximal quantum violation of $I_3^+$, the underlying quantum state must also have a negativity that approaches 1, which is exactly the negativity of $\ket{\Phi^+_3}$. However, the corresponding quantum correlation $\vec{P}^{I_3^+}_\Q$ --- as was shown in Ref.~\cite{Liang:PRA:2009} --- exhibits only a white-noise visibility $\vcr^{I^+_3}$ of 87.94\%.

In addition, while being a natural generalization of the CHSH Bell inequality to three ternary-outcome measurements, the Bell inequality $I^+_3$ is provably~\cite{Liang:PRA:2009} {\em not} facet-defining. Thus, it can be written as a convex combination of facets~\cite{Polytopes} such that at least one of which offers at least as good (if not better) white-noise visibility for the correlation $\vec{P}^{I_3^+}_\Q$. In particular, if this facet is also maximally violated by $\ket{\Phi^+_3}$, one will have obtained a more-promising candidate for the self-testing of a maximally entangled two-qutrit state.

How then do we look for the constituent facet-defining Bell inequalities which give rise to inequality $I_3^+$? It turns out that in computing the critical visibility $\vcr$ of any given nonlocal correlation $\vecP_\Q$, the (dual of the) linear program also outputs a Bell inequality $\I$ such that $\vecP_\Q$ has a visibility of $\vcr$ with respect to $\I$ (see Appendix~\ref{App:LP} for details). For a {\em generic} nonlocal correlation $\vecP_\Q$, as we decrease the weight $v$, the convex combination given in~\eqref{Eq:ConvexCombi} enters the local polytope $\L$ via one of its facets and the Bell inequality outputted by the linear program is thus also facet-defining.

Indeed, by solving the linear program of \eqref{Eq:LP:Visibility} using  the correlation $\vec{P}^{I_3^+}_\Q$,  we obtain the 16th facet-defining Bell inequality listed in table~\ref{Tab:BellCoeff} with the same critical visibility offered by $I_3^+$.  More generally, by maximizing the quantum violation of $I_3^+$ using two-qutrit entangled states of the form
\begin{equation}\label{eq:TwoQutritStateWithGamma}
	\ket{\psi(\gamma,\gamma')} = \frac{\ket{0}_\As\ket{0}_\Bs + \gamma \ket{1}_\As\ket{1}_\Bs 
	+ \gamma'\ket{2}_\As	\ket{2}_\Bs}{\sqrt{1+\gamma^2+\gamma'^2}},
\end{equation}
with $\gamma \in [0,1]$ and $\gamma'=1$, we obtain nonlocal correlations that exhibit $\vcr^{I^+_3}\in[0.8794, 0.9683]$.
Then, using these correlations as input to the linear program given in \eqref{Eq:LP:Visibility}, we obtain another 15 facet-defining Bell inequalities, listed as inequality 1 to 15 in table~\ref{Tab:BellCoeff}. Interestingly, among these 16 facets, only 12  are genuinely novel facets for the Bell scenario $\{[3\, 3\, 3]\, [3\, 3\, 3]\}$ while inequality 12, 13, 15, and 16 --- as can be verified using the online platform~\cite{faacet} (see also Ref.~\cite{Rosset:Faacet}) --- are reducible, respectively, to the simpler Bell scenario $\{[3\, 3\, 2]\, [3\, 3\, 2]\}$ and $\{[3\, 2]\, [2\, 2\, 2]\}$.\footnote{It turns out that the 16th inequality was already discovered in Ref.~\cite{Pironio:AllCHSH} and is the only non-CHSH-type facet for the scenario $\{[3\, 2]\, [2\, 2\, 2]\}$.} Finally, we find two other facets by using the (post-processed) experimentally-observed correlations (see section~\ref{Sec:DI}) as input to the linear program \eqref{Eq:LP:Visibility}. Hereafter, we denote all these inequalities by $\ImaxN{n}$ with $n\in{1,2,\ldots,18}$, i.e.,
\begin{equation}
	\ImaxN{n}: \S_n\equiv\vec{\beta}_n\cdot\vec{P} \stackrel{\L}{\le}  \S_{\tiny \mbox{max}}^\L({\vec{\beta}_n}).
\end{equation}

\subsection{Quantum violation of Bell inequalities}

Evidently, as can be seen in table~\ref{Tab:BellIneqProp}, all the 18 Bell inequalities presented can be violated quantum mechanically with a visibility $\vcr=v_{\mbox{\tiny max}}$ much better than the starting visibility $\vcr^{I^+_3}\in[0.8794,0.9683]$. Moreover, except for $\ImaxN{16}$ and $\ImaxN{17}$ --- both of which can be reduced to a simpler Bell scenario involving a combination of binary-outcome and ternary-outcome measurements --- the maximal quantum violation\footnote{These quantum violations were obtained using the optimization techniques presented in Refs.~\cite{yc-tools1,Liang:PRA:2009} and verified to be the quantum maximum using a convergent hierarchy of semidefinite programs (SDP) proposed by Navascu\'es-Pironio-Ac\'in (NPA)~\cite{hierarchy1} (see also Refs.~\cite{hierarchy2,Moroder:PRL:PPT}).} of the rest can already be achieved using two-qubit entangled states, including the maximally entangled two-qubit state $\ket{\Phi^+}=\ket{\psi(1,0)}$, cf. \eqref{eq:TwoQutritStateWithGamma}. These inequalities therefore serve, to our knowledge, the first examples of non-lifted facet-defining Bell inequalities whose maximal quantum violation is attainable using a Hilbert space dimension {\em smaller} than the number of possible outcomes involved.

\begin{table*}[h!p]
\begin{center}
\caption{\label{Tab:BellCoeff} Coefficients of  facet-defining Bell inequalities $\ImaxN{n}: \S_n=\vec{\beta}\cdot\vec{P}\le\S^\L_{\mbox{\tiny max}}$ found by solving the linear program given in \eqref{Eq:LP:Visibility}. The left-most column gives the inequality number, i.e., $n$ in $\ImaxN{n}$, whereas the second column gives the local bound $\S^\L_{\mbox{\tiny max}}$, i.e., the maximum value of the Bell expression allowed by all $\vecP\in\L$, cf. \eqref{Eq:CGForm}. The coefficients $\vec{\beta}$ of the Bell inequalities are sorted according to the order in which they appear in the table of \eqref{Eq:CGtable}, i.e., first from the top row to the bottom row, then from the left-most column to the right-most column: $\beta_{A,0}^0$, $\beta_{A,1}^0$, $\beta_{A,0}^1$, $\ldots$, $\beta^{22}_{11}$.
}
\footnotesize\setlength{\tabcolsep}{1.7pt}
\begin{tabular}{r|c|rrrrrrrrrrrrrrrrrrrrrrrrrrrrrrrrrrrrrrrrrrrrrrrr}
\hline\hline
$n$ & $\S^\L_{\mbox{\tiny max}}$ & \multicolumn{48}{c}{$\vec{\beta}_n$}\\ \hline
 1 &  2 &  0 &  1 &  0 &  1 & -1 &  0 &  0 &  1 &  0 &  1 & -2 &  1 & -1 &  1 &  0 &  0 &  0 & -1 &  0 & -1 & -1 &  1 &  0 &  0 &  1 &  0 & -1 &  0 & -2 & -1 &  1 &  0 & -1 &  1 &  0 &  1 &  0 & -1 & -1 &  0 &  1 &  1 & -1 & -1 & -1 &  0 &  1 & -1 \\
 2 &  2 & -1 &  1 &  0 &  1 & -1 &  0 &  0 &  2 &  0 &  1 & -2 &  1 & -2 &  1 &  1 &  0 &  0 & -1 &  0 & -1 & -1 &  1 &  0 &  0 &  1 &  0 &  0 &  0 & -1 & -1 &  1 &  0 & -1 &  1 &  0 &  1 &  0 & -1 & -1 &  0 &  1 &  1 & -1 & -1 & -1 &  0 &  1 & -1 \\
 3 &  2 &  0 &  1 &  0 &  1 & -1 &  0 &  0 &  1 &  0 &  1 & -2 &  1 & -2 &  1 &  0 &  0 &  0 & -1 &  0 & -1 & -1 &  1 &  0 &  0 &  1 &  0 & -1 &  0 & -1 & -1 &  1 &  0 &  0 &  1 &  0 &  1 &  0 & -1 & -1 &  0 &  1 &  1 & -1 & -1 & -2 &  0 &  1 & -1 \\
 4 &  2 &  0 &  1 & -1 &  0 &  0 &  1 & -1 &  2 &  1 &  1 & -1 &  1 & -2 &  1 &  0 &  0 &  0 & -1 &  0 & -1 &  0 &  1 &  0 &  0 &  1 & -1 & -2 &  1 & -1 & -1 &  1 &  0 & -1 &  0 & -1 &  1 &  0 &  0 &  0 &  0 &  1 &  0 & -2 & -1 &  0 &  1 &  1 & -1 \\
 5 &  2 &  0 &  1 &  0 &  1 & -1 &  0 &  0 &  1 &  0 &  1 & -2 &  1 & -2 &  1 &  0 &  0 &  0 & -1 &  0 & -1 & -1 &  1 &  0 &  0 &  1 &  0 &  0 &  0 & -1 & -1 &  1 &  0 &  0 &  1 &  0 &  1 &  0 & -1 & -2 &  0 &  1 &  1 & -1 & -1 & -2 &  0 &  1 & -1 \\
6 &  2 & -1 &  1 &  0 &  1 & -1 &  0 &  0 &  2 &  0 &  1 & -2 &  1 & -2 &  1 &  1 &  0 &  0 & -1 &  0 & -1 & -1 &  1 &  0 &  0 &  1 &  0 &  0 &  0 & -1 & -1 &  1 &  0 &  0 &  1 &  0 &  1 &  0 & -1 & -1 &  0 &  1 &  1 & -1 & -1 & -2 &  0 &  1 & -1 \\
 7 &  2 &  0 &  1 &  0 &  1 & -1 &  0 &  0 &  1 &  0 &  1 & -2 &  1 & -2 &  1 &  0 &  0 &  0 & -1 &  0 & -1 & -1 &  1 &  0 &  0 &  1 &  0 & -1 &  0 & -1 & -1 &  1 & -1 &  0 &  1 &  0 &  1 &  0 & -1 & -1 &  0 &  1 &  1 & -1 & -1 & -2 &  0 &  1 &  0 \\
8 &  2 & -1 &  1 & -1 &  0 &  0 &  1 &  0 &  2 &  0 &  1 & -2 &  1 & -2 &  1 &  1 &  0 &  0 & -1 &  0 & -1 &  0 &  1 &  0 &  0 &  1 & -1 & -1 &  1 & -1 & -1 &  1 &  0 & -1 &  0 & -1 &  1 &  0 &  0 &  0 &  0 &  1 &  0 & -1 & -1 & -1 &  1 &  1 & -1 \\
 9 &  2 &  0 &  1 & -1 &  0 &  0 &  1 &  0 &  1 &  0 &  1 & -2 &  1 & -2 &  1 &  0 &  0 &  0 & -1 &  0 & -1 &  0 &  1 &  0 &  0 &  1 & -1 & -2 &  1 & -1 & -1 &  1 &  0 & -2 &  0 & -1 &  1 &  0 &  0 &  0 &  0 &  1 &  0 & -1 & -1 &  0 &  1 &  1 & -1 \\
10 &  1 &  0 &  1 & -1 &  0 & -1 &  0 &  0 &  1 &  0 &  1 & -2 &  1 & -1 &  1 &  0 & -1 &  0 & -1 &  0 & -1 & -1 &  1 &  0 &  0 &  1 &  0 &  0 &  0 & -2 & -1 &  1 & -1 &  0 &  1 & -1 &  1 &  0 &  0 & -1 &  0 &  1 &  0 & -1 & -1 & -1 &  1 &  1 &  0 \\
11 &  2 &  0 &  1 &  0 &  1 & -1 &  0 &  0 &  1 &  0 &  1 & -2 &  1 & -1 &  1 &  0 &  0 &  0 & -1 &  0 & -1 & -1 &  1 &  0 &  0 &  1 &  0 & -1 &  0 & -2 & -1 &  1 &  0 &  0 &  1 &  0 &  1 &  0 & -1 & -1 &  0 &  1 &  1 & -1 & -1 & -2 &  0 &  1 & -1 \\
12 &  2 &  0 &  1 &  0 &  1 & -1 &  0 &  1 &  1 &  0 &  0 & -2 &  1 & -1 &  1 &  1 &  0 &  0 & -1 &  0 & -1 &  0 &  0 &  0 &  0 &  0 &  0 &  0 &  1 & -2 & -1 &  0 & -1 & -1 &  1 & -1 &  1 &  0 &  0 &  0 &  0 &  1 &  0 & -1 & -1 &  0 &  1 &  1 & -1 \\
13 &  2 & -1 &  1 & -1 &  0 &  0 &  1 &  0 &  2 &  0 &  1 & -2 &  1 & -2 &  1 &  1 &  0 &  0 & -1 &  0 & -1 &  0 &  1 &  0 &  0 &  1 & -1 & -1 &  1 & -1 & -1 &  1 &  0 & -2 &  0 & -1 &  1 &  0 &  0 &  0 &  0 &  1 &  0 & -1 & -1 &  0 &  1 &  1 & -1 \\
14 &  2 &  0 &  1 &  0 &  1 & -1 &  0 &  0 &  1 &  0 &  1 & -2 &  1 & -1 &  1 &  0 &  0 &  0 & -1 &  0 & -1 & -1 &  1 &  0 &  0 &  1 &  0 &  0 &  0 & -2 & -1 &  1 &  0 &  0 &  1 &  0 &  1 &  0 & -1 & -2 &  0 &  1 &  1 & -1 & -1 & -2 &  0 &  1 & -1 \\
15 &  1 & -2 &  1 &  0 &  0 &  0 & -1 &  0 &  2 & -1 &  0 & -1 &  1 & -1 &  1 &  1 & -1 &  0 & -1 &  0 & -1 & -1 &  1 & -1 &  0 &  2 & -1 &  1 &  0 &  0 & -1 &  0 &  1 &  0 &  1 &  0 &  1 &  0 &  0 & -1 & -1 &  1 &  0 &  0 &  0 &  0 &  0 &  0 &  0 \\ \hline
16 &  1 &  0 &  1 &  0 &  0 &  0 &  0 & -1 &  1 &  0 &  1 &  0 &  0 &  0 &  0 &  0 &  0 &  0 &  0 &  0 &  0 &  0 &  0 &  0 &  0 &  0 &  0 &  0 &  0 & -1 & -1 &  1 &  0 &  0 &  0 &  1 &  0 & -1 & -1 &  0 &  0 &  0 &  1 &  0 & -1 & -1 &  0 &  0 &  0 \\ \hline
17 &  1 &  1 &  1 &  0 &  0 &  0 & -1 &  0 &  0 &  0 &  0 &  0 &  0 &  0 &  1 & -1 & -1 & -1 &  0 & -1 & -1 &  0 &  0 &  0 & -1 &  0 &  0 &  1 &  0 & -1 & -1 & -1 &  0 &  1 &  1 &  0 &  0 &  0 &  0 &  0 &  0 &  0 &  0 & -1 & -1 &  1 &  0 &  0 &  1 \\
18 &  1 &  1 &  1 &  0 &  0 &  0 & -1 &  0 &  0 &  0 &  0 &  0 &  0 &  0 &  1 & -1 & -1 &  0 &  0 & -1 & -1 &  0 &  0 &  0 & -1 &  0 &  0 &  1 &  0 & -1 & -1 & -1 &  0 &  1 &  1 &  0 &  0 &  0 &  0 &  0 &  0 &  0 & -1 & -1 & -1 &  1 &  0 &  1 &  2 \\ \hline
19 & 1  &  0 &  1 &  0  &  0 &  0 &  0 &  0 &  0 & -1 & 0 & -1 & 1 &  0 & 1 & -1 & -1 & -1 & -1 &  0 &  -1 &  0 &  0 & -1 & 1 & 0 & -1 &  0 &  0 & -1 &  -1 & 0 & -1 &  0 &    1 &    0 &  1     & 0 & -1 & 0 & -1&  0 &  0 &  0 & -1 &  0 & 1 & 0 &  0\\
\hline\hline
\end{tabular}
\end{center}
\end{table*}

\begin{table*}[h!p]
\caption{\label{Tab:BellIneqProp} Summary of some of the properties of the Bell inequalities $\ImaxN{n}: \S_n\le\S^\L_{\mbox{\tiny max}}$ (3rd-7th column) and $\IminN{n}: \S_n\ge\S^\L_{\mbox{\tiny min}}$ (8rd-12th column) for the Bell expression $\S_n$ presented in table~\ref{Tab:BellCoeff}. The second column gives the simplest Bell scenario to which $\S_n$ can be reduced. Within each block column, we list  the maximum (minimum) local value $\S^\L_{\mbox{\tiny max}}$ ($\S^\L_{\mbox{\tiny min}}$), the maximum (minimum) quantum value $\S^\Q_{\mbox{\tiny max}}$, the (Schmidt coefficients of the) state $\ket{\psi_{\mbox{\tiny max}}}$  ($\ket{\psi_{\mbox{\tiny min}}}$) found to achieve $\S^\Q_{\mbox{\tiny max}}$ ($\S^\Q_{\mbox{\tiny min}}$), the corresponding state visibility $\nucr^I$ and white-noise visibility $\vcr^I$. The  dimension spanned by  the set of local extreme points saturating the minimal local value $\S^\L_{\mbox{\tiny min}}$ is given in the last column. Entries marked with $^\dag$ means that within the numerical precision of the solver, the corresponding Bell inequality $\IminN{n}$ is provably satisfied by quantum theory. For the entry marked with $\ddag$, there remains a gap of the order of $2.7\times10^{-3}$ between the $\S^\Q_{\mbox{\tiny max}}$ presented and the best upper bound that we obtained from local level 2 of the hierarchy considered in Ref.~\cite{Moroder:PRL:PPT}.
}
\begin{tabular}{cc|ccccc|ccccccc}
\hline\hline
$n$   & Scenario  &$\S^\L_{\mbox{\tiny max}}$ & $\S^\Q_{\mbox{\tiny max}}$  & $\ket{\psi_{\mbox{\tiny max}}}$ & $\nu_{\mbox{\tiny max}}$ & $v_{\mbox{\tiny max}}$ 
& $\S^\L_{\mbox{\tiny min}}$ & $\S^\Q_{\mbox{\tiny min}}$ & $\ket{\psi_{\mbox{\tiny min}}}$ & $\nu_{\mbox{\tiny min}}$  & $v_{\mbox{\tiny min}}$ & $d_{min}$\\ 
\hline\hline
1 & $\{[3\, 3\, 3]\, [3\, 3\, 3]\}$ & 2 & 2.6972 & $\ket{\Phi^+}$ &  0.7819 & 0.7415  & -4$^\dag$ & -4.0000 & - & - & - & 1\\
2 & $\{[3\, 3\, 3]\, [3\, 3\, 3]\}$ & 2 & 2.6712& $\ket{\Phi^+}$ &  0.7884 & 0.7487  & -4$^\dag$ & -4.0000 & -& -  & -& 1\\
3 & $\{[3\, 3\, 3]\, [3\, 3\, 3]\}$ & 2 & 2.6586  & [0.7278; 0.6857]  & 0.7915 & 0.7523  & -5$^\dag$ & -5.0000 & - & - & - & 2\\
4 & $\{[3\, 3\, 3]\, [3\, 3\, 3]\}$ & 2 & 2.6586  & [0.7278; 0.6857] &  0.7915 & 0.7523  & -5$^\dag$ & -5.0000 & - & - & - & 2\\
5 & $\{[3\, 3\, 3]\, [3\, 3\, 3]\}$ & 2 & 2.6488  & [0.7129; 0.7013] & 0.7940 & 0.7551   & -4$^\dag$ & -4.0000 & - & - & - & 4\\
6 & $\{[3\, 3\, 3]\, [3\, 3\, 3]\}$ & 2 & 2.6577  & [0.7222; 0.6917] &  0.7917 & 0.7525  & -4$^\dag$ & -4.0000 & - & - & - & 4\\
7 & $\{[3\, 3\, 3]\, [3\, 3\, 3]\}$ & 2 & 2.6577  & [0.7222; 0.6917] & 0.7917 & 0.7525  & -4$^\dag$ & -4.0000 & - & - & - & 5\\
8 & $\{[3\, 3\, 3]\, [3\, 3\, 3]\}$ & 2 & 2.6577  & [0.7222; 0.6917] &  0.7917 & 0.7525  & -4 & -4.0010 & [0.7925; 0.5893; 0.1568] & 0.9997 & 0.9997 & 5\\
9 & $\{[3\, 3\, 3]\, [3\, 3\, 3]\}$ & 2 & 2.6720 & [0.7150; 0.6992] &  0.7881 & 0.7485  & -4 & -4.0171 & [0.7468; 0.6175; 0.2470]  &  0.9941 & 0.9941 & 6\\
10 & $\{[3\, 3\, 3]\, [3\, 3\, 3]\}$ & 1 & 1.6720  & [0.7150; 0.6992] &  0.7881 & 0.7485  & -5 & -5.0171 & [0.7468; 0.6175; 0.2470]  & 0.9941 & 0.9941& 7\\
11 & $\{[3\, 3\, 3]\, [3\, 3\, 3]\}$ & 2 & 2.6955  & [0.7230; 0.6908] &  0.7824 & 0.7420  & -4 & -4.0138 & [0.7932; 0.5571; 0.2457] & 0.9952 & 0.9952 & 8\\
12 & $\{[3\, 3\, 2]\, [3\, 3\, 2]\}$ & 2 & 2.5820  & [0.7258; 0.6879] &  0.7746 & 0.7277  &-3 &  -3.0005 & [0.8957; 0.4328; 0.1023] & 0.9998 & 0.9998 & 8\\
13 & $\{[3\, 3\, 2]\, [3\, 3\, 2]\}$ & 2 & 2.6712 & $\ket{\Phi^+}$ &  0.7884 & 0.7487  & -3 & -3.6712 & $\ket{\Phi^+}$ & 0.7884 & 0.8172  & 26\\
14 & $\{[3\, 3\, 3]\, [3\, 3\, 3]\}$ & 2 & 2.6972  & $\ket{\Phi^+}$ &  0.7819 & 0.7415  & -3 & -3.6972 & $\ket{\Phi^+}$&  0.7819 & 0.8114  & 27\\
15 & $\{[3\, 3\, 2]\, [3\, 3\, 2]\}$ & 1 & 1.5923 & $\ket{\Phi^+}$ &  0.7715 & 0.7242  & -3 & -3.5923 & $\ket{\Phi^+}$ & 0.7715 & 0.8050 & 29\\
16 & $\{[3\, 2]\, [2\, 2\, 2]\}$ & 1 & 1.2532  & [0.6608; 0.5307; 0.5307] &  0.7247 & 0.7247  & -1 & -1.0328 & [0.6773; 0.6769; 0.2882] & 0.9682 & 0.9682 & 30\\
\hline
17 & $\{[3\, 2\, 2]\, [3\, 2\, 2]\}$ & 1 & 1.3090  & [0.6388; 0.5860; 0.4985] & 0.7639 & 0.7639  & -3$^\dag$ &  -3.0000 & - & - & - & 4\\
18 & $\{[3\, 2\, 2]\, [3\, 2\, 2]\}$ & 1 & 1.4142 & $\ket{\Phi^+}$ &   0.7071 &   0.7071 & -2$^\dag$ &  -2.0000 & - & - & - & 16\\ \hline
19 & $\{[3\, 3\, 3]\, [3\, 3\, 3]\}$ & 1 & 1.3782$^\ddag$  &  [0.6069; 0.6064; 0.5137] &    0.7925 & 0.7925  &-3 & -3.2071 & $\ket{\Phi^+}$ &  0.7071 & 0.9250 & 13\\
\end{tabular}
\end{table*}

Given the above observation, a natural question that one may now ask is whether a genuine POVM (i.e., measurement involving non-projective operators) is ``needed" to achieve these maximal quantum violations. Of course, if there is no restriction in the Hilbert space dimension, the maximal quantum violation of a Bell inequality can always be achieved by performing only projective measurements in a sufficiently large Hilbert space (see, for instance, Refs.~\cite{MikeIke:Book,Brandt:JOB:2003}). The above question is thus relevant only if we restrict ourselves to the minimal Hilbert space dimension where the maximal quantum violation of a given Bell inequality is known to be achievable. 

Interestingly, we have found that for all but one of these ``qubit" inequalities, it is already sufficient to allow the trivial projector $\mathbf{0}_2$, i.e., the $2\times 2$ zero matrix in the three-outcome measurement in order to recover the maximal quantum violation.  The only exception to these is $\ImaxN{12}$, where we could show --- by considering all possible combinations of trivial-projector assignments and using a suitable modification of the SDP of Ref.~\cite{hierarchy1} as well as the SDP discussed in Ref.~\cite{yc-tools1} --- that to achieve maximal quantum violation of $\ImaxN{12}$ using a quantum state of the lowest possible dimension necessarily requires non-projective measurements.  In contrast with previous example~\cite{Vertesi:POVM} where a genuine POVM was found to be relevant, our example here actually makes use of a facet-defining Bell inequality, thus answering a stronger form of the opened problem (fundamental question no.~10) posed by Gisin in Ref.~\cite{Gisin:Problems}. Moreover, our example is also considerably more noise-resistant than that of Ref.~\cite{Vertesi:POVM}, as we can tolerate between 6.66\%--17.54\%\footnote{The uncertainty in this critical value of white noise stems from the gap between the best lower bound on quantum violation that we could find and the best upper bound that we could obtain assuming projective measurements.} of white noise, before losing  the advantage of genuine POVM over projective measurements using $\ImaxN{12}$. For an explicit set of genuine POVM leading to the maximal quantum violation of $\ImaxN{12}$, see Appendix~\ref{App:OptMeas12}.

For completeness, we also analyze the quantum violation of the only  $\{[3\, 3\, 3]\, [3\, 3\, 3]\}$  facet known in the literaute~\cite{Collins:JPA:2004}, which we list as $\ImaxN{19}$ (and cast in a more symmetrical form) in table~\ref{Tab:BellCoeff} . Likewise, we also carry out detailed analysis of the Bell inequality that result from the {\em minimum} [cf.~\eqref{Eq:CGForm}] of the Bell expression given in table~\ref{Tab:BellCoeff} over all correlations in $\L$, i.e., 
\begin{equation}\label{Eq:CGForm:Min}
	\I^{\mbox{\tiny min}}_n: S_n=\vec{\beta}_n\cdot\vec{P} \stackrel{\L}{\ge}  \S_{\tiny \mbox{min}}^\L({\vec{\beta}_n}).
\end{equation}
The results of these investigations are also summarized in table~\ref{Tab:BellIneqProp}. Not surprisingly, all the Bell inequalities $\I^{\mbox{\tiny min}}_n$ with $n\in\{1,2,\ldots,19\}$ are {\em not} facet-defining. Moreover, from the numerical upper bound obtained from Ref.~\cite{hierarchy1}, 9 of these inequalities cannot even be violated quantum mechanically, while 5 others can be violated  with rather poor visibility $\vcr^I$ using some partially entangled two-qutrit states. The remaining 4 inequalities $\IminN{13}$, $\IminN{14}$, $\IminN{15}$, and $\IminN{19}$ turn out to be maximally violated by $\ket{\Phi^+}$. Moreover, the maximal quantum violation of $\IminN{n}$ with $n\in\{13, 14, 15\}$ turns out to give exactly the same state visibility as their counterpart $\ImaxN{n}$.

For $\ImaxN{19}$, the best quantum violation that we have found is attained using a partially entangled two-qutrit state. However, since we cannot close the gap between this quantum violation with the upper bound coming from the NPA hierarchy~\cite{hierarchy1} (or the hierarchy considered in Ref.~\cite{Moroder:PRL:PPT}), we cannot yet conclude that the presented value indeed represents the maximal quantum violation of $\ImaxN{19}$. Finally, let us note from table~\ref{Tab:BellIneqProp} that the Bell inequalities listed within each of the three sets $\{\ImaxN{3}, \ImaxN{4}\}$, $\{\ImaxN{6}, \ImaxN{7}, \ImaxN{8}\}$, $\{\ImaxN{9}, \ImaxN{10}\}$ share surprisingly many common properties. In fact, it is impossible to know that inequalities $\ImaxN{3}$ and $\ImaxN{4}$ are inequivalent~\cite{Collins:JPA:2004} [i.e., cannot be transformed from one to the other via~\eqref{Eq:NS} and/or relabeling of parties, inputs and outputs] based on the properties that we have analyzed. Their inequivalence is eventually confirmed by computing their canonical representation~\cite{Rosset:Faacet} via the platform provided in Ref.~\cite{faacet}.

\section{Experimental Demonstration of Bell Inequality Violation}
\label{sec:ExpDemo}

In this section, we present how energy-time entangled photon pairs can be realized experimentally in order to demonstrate the quantum violation of Bell inequality $I_3^+$ [\eqref{Ineq:I3plus}] using a one-parameter family of entangled two-qutrit states, i.e.,~\eqref{eq:TwoQutritStateWithGamma} with $\gamma'=1$, as well as the violation of $\ImaxN{12}$ and $\ImaxN{14}$ (table~\ref{Tab:BellCoeff}) using a one-parameter family of entangled two-qubit states, i.e.,~\eqref{eq:TwoQutritStateWithGamma} with $\gamma'=0$. Note that $I_3^+$ is the only known Bell inequality with maximal quantum violation achieved by a maximally entangled two-qutrit state, while $\ImaxN{14}$ is a ternary-outcome Bell inequality that is maximally violated by a maximally entangled two-qubit state, whereas $\ImaxN{12}$ is maximally violated by performing a genuine POVM, or in other words, {\em non-projective measurements} on a partially entangled two-qubit state.

\subsection{Experimental Setup}
\label{subsec:ExpSetup}
\begin{figure*}[t]
\begin{minipage}{0.5\textwidth}
\centerline{\epsfig{file=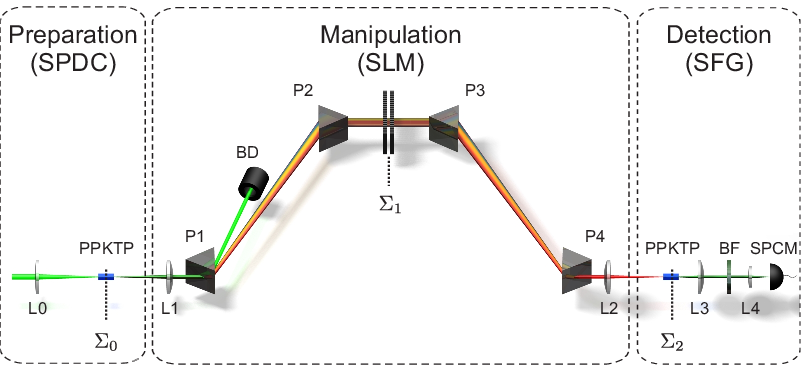,width=8.5cm}}
\end{minipage}
\begin{minipage}{0.4\textwidth}
\centerline{\epsfig{file=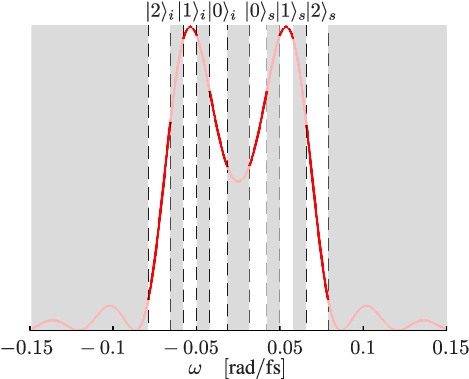,width=5.75cm}}
\end{minipage}
\caption{\textit{Left}: Schematic overview of the experimental setup. \textit{Preparation}: Pump beam is focused (L0, $f = 150$ mm) into a nonlinear PPKTP crystal for SPDC.  \textit{Manipulation}: The pulse shaper consists of a symmetric two-lens imaging arrangement (L1, L2, $f = 100$ mm) and a four-prism  (P1-P4) compressor. At its symmetry axis an  SLM is located and used in transmission mode. The residue of the pump beam is chocked off by a beam dump BD. \textit{Detection}: Identical nonlinear crystal for SFG used such that after a two-lens imaging system (L3, L4 , $f = 60$ mm, $f = 11$ mm) the up-converted photons are detected via a SPCM. The entangled photons pass through the imaging sequence $\Sigma_0\rightarrow\Sigma_1\rightarrow\Sigma_2$ with corresponding magnification factors $1:6:1$ and the bandpass filter BF is used to filter out the non-up-converted photons. \textit{Right:} Simulated SPDC spectrum overlaid with a schematic frequency-bin pattern. By means of the SLM, the transmitted amplitude $\vert u_j^{i,s}\vert$ (white bars) and phase $\phi_j^{i,s}$ of each bin can be controlled independently according to \eqref{eq:Mslm}. }
\label{fig:expSetupAndFrequencyBins}
\end{figure*}

As depicted in figure~\ref{fig:expSetupAndFrequencyBins}, our experimental setup \cite{bernhard2014, schwarz2014} can be subdivided into three parts: preparation, manipulation and detection of entangled photons. By pumping a periodically poled KTiOPO$_4$ (PPKTP) crystal with a quasi-monochromatic ND:YVO$_4$ (Coherent Verdi V5) laser with a central wavelength $\lambda_{p,c}=532$ nm, collinearly phase-matched type-0 spontaneous parametric down-conversion (SPDC) is exploited. Up to first-order in perturbation theory, the corresponding biphoton state is described by
\begin{equation}
\vert \psi \rangle = \int_{-\infty}^{\infty} d\omega \, \Lambda(\omega) \, \hat{a}_i^{\dag}(\omega) \, \hat{a}_s^{\dag}(-\omega)\vert 0\rangle_i\vert 0\rangle_s,
\label{eq:biphtonStateDC}
\end{equation} 
where the leading order vacuum state is omitted. By acting on the composite vacuum state $\vert 0 \rangle_i \vert 0 \rangle_s$, the operators $\hat{a}_{i,s}^{\dag}(\omega)$ create the idler ($i$) and signal ($s$) photon at relative frequency $\omega$ (with respect to $\frac{\omega_{p,c}}{2}$). Since we assume a continuous pump field as well as degenerate center frequencies $\omega_{i,c} = \omega_{s,c} = \tfrac{\omega_{p,c}}{2}$, for idler and signal photon, respectively, the joint spectral amplitude (JSA), in general being a two-dimensional function denoted by $\Lambda(\omega_i, \omega_s)$, simplifies to $\Lambda(\omega)$ with $\omega \equiv \omega_s = -\omega_i$, as used in \eqref{eq:biphtonStateDC}.

The subsequent manipulation part is realized via a prism-based pulse shaping configuration including a spatial light modulator (SLM, Jenoptik, SLM-S640d) as a reconfigurable modulation device. The SLM is endowed with two nematic liquid crystal displays and is used in transmission mode. The respective effect on each photon spectrum can be described by a complex transfer function $M^{i,s}(\omega)$ which transforms the JSA according to
\begin{equation}
\tilde{\Lambda}(\omega) = \Lambda(\omega)M(\omega),
\label{eq:JSAmodbySLM}
\end{equation}
where $M(\omega) = M^i(\omega)M^s(-\omega)$. Finally, coincidences between entangled photons are measured by sum-frequency generation (SFG) using a second PPKTP crystal. The resulting up-converted photons are then imaged onto the photosensitive area of a single photon counting module (SPCM, ID Quantique id100-20-uln). Accordingly, the signal detected by the latter is described by the first-order coherence function 
\begin{equation}
S \propto \left\vert \int_{-\infty}^{\infty} d\omega \, M(\omega) \Gamma(\omega) \right\vert^2,
\label{eq:expSignal}
\end{equation}
where the joint spectral amplitude is modified by the acceptance bandwidth of the detection crystal $\Phi(\omega)$ such that $\Gamma(\omega) \propto \Lambda(\omega)\Phi(\omega)$. To incorporate the finite spectral resolution at the SLM plane we further convolute the JSA with a Gaussian modelled point-spread function (PSF) denoted with $\Upsilon_{PSF}(\omega)$ according to 
\begin{equation}
\Gamma(\omega) \rightarrow \Gamma_{PSF}(\omega) \propto (\Gamma \otimes \Upsilon_{PSF})(\omega).
\end{equation}

\subsection{Projective Measurements in a Frequency-Bin Basis}
\label{Sec:Exp:Meas}

\begin{figure*}[ht!]
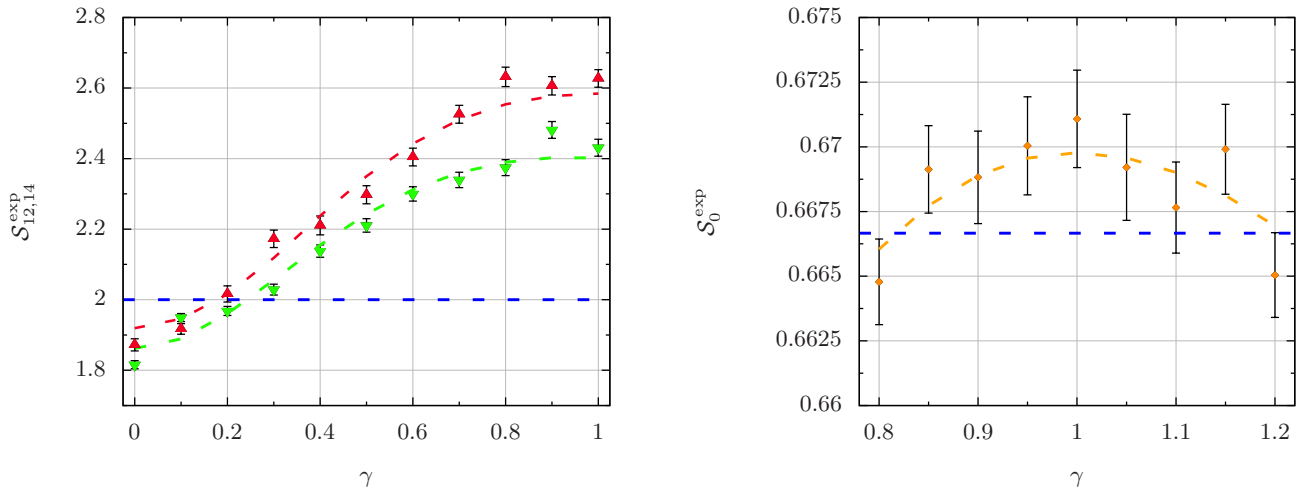

  \captionsetup[subfigure]{labelformat=empty}
  \subfloat[]{\resizebox{0.49\textwidth}{!}{\input{I12I14_Qubit}}}\quad
  \subfloat[]{\resizebox{0.49\textwidth}{!}{\input{I3_Qutrit_detail}}}
  \caption{Experimentally measured Bell parameters $\SexpI{n}$ with $n = \{0,12,14 \}$ shown as a function of the entanglement parameter $\gamma$ with $1\sigma$ uncertainty. \textit{Left:} Experimentally measured Bell parameter $\SexpI{n}$ for the Bell inequality $\ImaxN{12}$ (green downturned triangles) and $\ImaxN{14}$ (red upturned triangles) using a two-qubit state $\vert \psi(\gamma,0)\rangle$ with $\gamma \in [0,1]$ and step-size $\Delta\gamma=0.1$. The average state visibility $\nu_d^I$, cf. \eqref{Eq:StateVisibility}, for $\ImaxN{12}$ is  $\nu^{\mathcal{I}_{12}}_{3} = 0.9182 \pm 0.0073$ whereas for $\ImaxN{14}$ we obtain  $\nu^{\mathcal{I}_{14}}_{2} = 0.9561 \pm 0.0089$.
\textit{Right:} Experimentally measured Bell parameter $\SexpI{0}$ for the Bell inequality $I^+_3$ using two-qutrit state $\vert \psi(\gamma,1)\rangle$ with $\gamma \in [0.8,1.2]$ and step-size $\Delta\gamma=0.05$. The average state visibility is $\nu_3^{I^+_3} = 0.8876 \pm 0.0032$. In both plots, the (horizontal) blue dashed line indicates the local bound $\S_n=\Smax^\L(\vec{\beta}_n)$ whereas the coloured dashed lines show the theoretical predictions, scaled with the corresponding state visibility.}
  \label{fig:resultsI12I14QubitAndI3Qutrit}
\end{figure*}

To access entangled two-qudit states we project the continuous biphoton state of \eqref{eq:biphtonStateDC} onto a $d^2$-dimensional discrete subspace spanned by orthonormal basis states $\vert j\rangle_i \vert k\rangle_s$, where $\vert j\rangle_{i,s} \equiv \int_{-\infty}^{\infty} f_j^{i,s}(\omega)\,\hat{a}_{i,s}^{\dag}(\omega)\vert 0\rangle_{i,s}$ (see Ref.~\citep{bessire2014} for details).
Here, the corresponding basis functions $f^{i,s}_j(\omega)$ are chosen according to a frequency-bin pattern which is given by
\begin{equation}
f_j^{i,s}(\omega) = \left\{
\begin{array}{cl} 
1/\sqrt{\Delta\omega_j} & \quad\text{for} \quad \vert \omega-\omega_j\vert < \Delta\omega_j/2, \\
0                     & \quad\text{otherwise}.
\end{array}
\right.
\label{eq:freqBins}
\end{equation}
If we impose that adjacent bins do not overlap, the (normalized) projected state can be rewritten in the form of a discrete two-qudit state
\begin{equation}
\vert \psi_d \rangle = \sum^{d-1}_{j=0}\,c_{j}\,\vert j\rangle_i \vert j \rangle_s.  
\label{eq:psi_diag}
\end{equation}
For $d=3$ the corresponding frequency-bin pattern, together with the associated basis states, is shown exemplarily in figure~\ref{fig:expSetupAndFrequencyBins}. We further, decompose the transfer functions $M^{i,s}(\omega)$ in terms of the basis functions $f_j^{i,s}(\omega)$ such that
\begin{equation}
M^{i,s}(\omega) = \sum_{j=0}^{d-1}\,u_j^{i,s}\,f_j^{i,s}(\omega) = \sum_{j=0}^{d-1}\,\vert u_j^{i,s}\vert\,e^{i\,\phi_j^{i,s}}\,f_j^{i,s}(\omega).
\label{eq:Mslm}
\end{equation}
Experimentally, the SLM consists of two liquid crystal displays. This allows us to manipulate the spectral phase $\phi_j^{i,s}$ and amplitude $\vert u_j^{i,s} \vert$ of the signal and idler photon independently \cite{baenz2014}. The latter is realized by an SLM induced polarization modulation on the entangled photons and an SFG crystal which only up-converts h-polarized photons. The combination of the SLM and the SFG coincidence detection finally realizes a projective measurement of $\vert \psi_d\rangle$ onto a direct product state $\vert \chi\rangle$ to be defined below. Accordingly, the discretized expression of \eqref{eq:expSignal} reads
\begin{equation}
S^{(d)} \propto \left\vert \sum_{j=0}^{d-1} u^i_j u^s_j c_j \right\vert^2 = \left\vert \langle \chi \vert \psi_d \rangle \right\vert^2.
\label{eq:S_proj}
\end{equation}
To be consistent with the framework of a Bell scenario we identify from now on the idler photon with Alice $(A \leftrightarrow i)$ and the signal photon with Bob $(B \leftrightarrow s)$. The set of correlations $P(a,b \vert x,y)$ is then related to $S^{(d)}$ according to $P(a,b \vert x,y)\propto S^{(d)}$ with a state $\vert \chi \rangle$ given by 
\begin{align}\label{Eq:chistate}
\vert \chi_{a\vert x,b\vert y} \rangle &= \vert a \rangle_A^x \vert b \rangle_B^y\\
 &= \left( \sum_{j=0}^{d-1} u^x_j(\{\eta_k^x\}) \vert j \rangle_A\right)\left( \sum_{j'=0}^{d-1} u^y_{j'}(\{\theta_k^y\}) \vert j' \rangle_B\right),\nonumber
\end{align}
where $M_{a|x}\otimes M_{b|y} = \vert a \rangle_A^x {^x_A\langle} a \vert \otimes \vert b \rangle_B^y {^y_B\langle} b \vert$. Each set of states $\{\ket{a}^x_A\}$, $\{\ket{b}^y_B\}$ in \eqref{Eq:chistate} represents the most general $d$-dimensional orthonormal basis according to SU($d$) and the optimal measurement settings $\lbrace\eta^x_k\rbrace$ and $\lbrace\theta^y_k\rbrace$ with $k \in \{1, ..., d^2-1\}$ are obtained by maximizing the violation of the Bell inequality under consideration. Experimentally,  the set of correlations $\Pexp$ is determined by measuring an average count rate $N\propto S^{(d)}$ over a certain time $T$. In particular, under the assumption that the source and measurements are both independent and identically distributed (i.i.d.), the observed correlation components $P_{\mbox{\tiny exp}}(a,b|x,y)$ are finally estimated via the relative frequencies
\begin{equation}\label{Eq:RawCorrelation}
	P_{\mbox{\tiny exp}}(a,b|x,y)=\frac{N(a,b,x,y)}{\sum_{a',b'} N(a',b',x,y)}
\end{equation}
for all $a,b,x,y$.

\subsection{Experimental Results}

In this section, experimental results obtained for the Bell inequality violation of $I_3^+$, $\ImaxN{12}$ and $\ImaxN{14}$ are provided. The quantum value for each of these inequalities is calculated using a representation of the corresponding Bell inequality expressed in the form of \eqref{Eq:betaS}. Specifically, to obtain the Bell coefficients $\alpha^{xy}_{ab}$ of $\ImaxN{12}$ and $\ImaxN{14}$ from the coefficients  given in table \ref{Tab:BellCoeff}, we apply the following transformation
\begin{align}
	\alpha^{xy}_{ab}&=\beta^{xy}_{ab}\quad\text{if}\,a,b\in\{0,1,2\}\,\, \text{and}\,\,\, x, y\in\{1,2\},\nonumber\\
	\alpha^{00}_{ab}&=\beta^{00}_{ab}+ \beta^0_{A,a}+\beta^0_{B,b}\quad \text{if}\, a, b\in\{0,1,2\},\\	
	\alpha^{x0}_{ab}&=\beta^{x0}_{ab}+ \beta^x_{A,a}\quad \text{if}\,a,b\in\{0,1,2\}\,\, \text{and}\,\,\, x\in\{1,2\},\nonumber\\
	\alpha^{0y}_{ab}&=\beta^{0y}_{ab}+ \beta^y_{B,b}\quad \text{if}\,a,b\in\{0,1,2\}\,\, \text{and}\,\,\, y\in\{1,2\},\nonumber
\end{align}
where all coefficients that were not defined in table~\ref{Tab:BellCoeff}, such as $\beta^{xy}_{22}$, $\beta^x_{A,2}$, $\beta^y_{B,2}$ etc., are understood to be zero in the above equation.

The experimental results ---  obtained by performing the optimized measurements for each inequality and for each fixed value of $\gamma$ --- are shown in figure~\ref{fig:resultsI12I14QubitAndI3Qutrit} as a function of $\gamma$. Note that for the experiment on $I_3^+$ and $\ImaxN{14}$, the preparation of a maximally entangled two-qutrit state $\ket{\Phi^+_3}$ and two-qubit state $\ket{\Phi^+}$, respectively, are of particular interest. The characteristic shape of a SPDC spectrum (figure~\ref{fig:expSetupAndFrequencyBins}), however, naturally leads to a non-maximally entangled quantum state, due to unequally distributed probability amplitudes. In our experiments, $\ket{\Phi^+}$ and  $\ket{\Phi^+_3}$ are thus prepared by applying the  Procrustean method of entanglement concentration \cite{bennett1996}. Moreover, for the quantum violation of $\ImaxN{12}$, in order to implement the non-projective POVM element, the two-qubit state $\ket{\psi(\gamma,0)}$ is embedded in a two-qutrit space, i.e., after preparing the state $\ket{\Phi^+_3}$ as described above, the two Schmidt coefficients $c_0$ and $c_1$ are changed according to table \ref{Tab:BellIneqProp} while setting $c_2$ to zero.

To quantify the imperfection in our setup, we consider the symmetric noise model of \eqref{Eq:StateVisibility} with $\rho$ being the pure state 
$\ket{\psi(\gamma,0)}$ or $\ket{\psi(\gamma,1)}$ of \eqref{eq:TwoQutritStateWithGamma} with the appropriate value of $\gamma$. From the measured quantum value of each inequality and the theoretically computed maximum quantum violation for each of the aforementioned states, we can then determine the mixing parameter $\nu_d^{\mathcal{I}}(\gamma)\in[0,1]$ that gives rise to the observed correlations. The final mixing parameter reported in the caption of figure~\ref{fig:resultsI12I14QubitAndI3Qutrit} is the mean value obtained by averaging $\nu_d^{\mathcal{I}}(\gamma)$ over different values of $\gamma$.

\begin{figure*}[t]
\centering
\resizebox{0.75\textwidth}{!}{\input{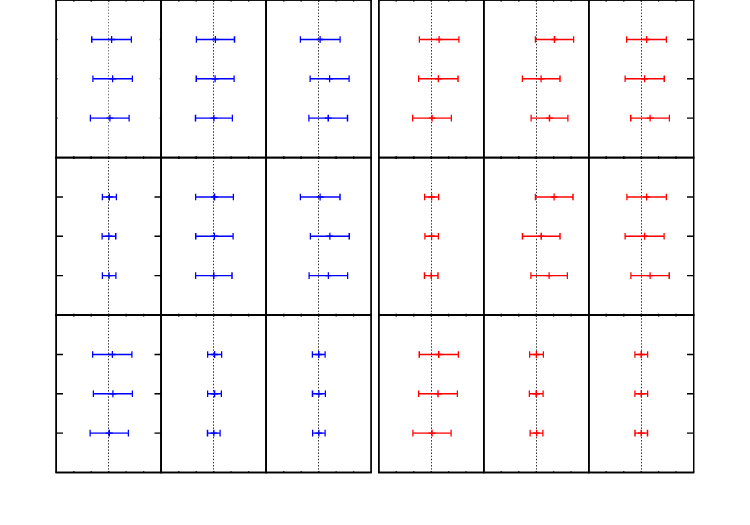}}
\caption{Differences in marginal probability distributions $\Delta P(a|x, y_{1,2})$ for Alice (blue plots on the left) and $\Delta P(b\vert x_{1,2}, y)$ for Bob (red plots on the right), evaluated for $\Pexp$ obtained from the measurement of the Bell inequality $\ImaxN{12}$ using qubits, with $\gamma = 1$ ($1\sigma$ uncertainties). The number pair $y_{1,2},x_{1,2} \in \{0,1,2\}$ refers to the pair of settings for which the difference according to \eqref{eq:NSinExpData_ax} and \eqref{eq:NSinExpData_by}, respectively, is evaluated.}
\label{fig:NonSignalingPlot_I12}
\end{figure*}

\section{Source characterization from outcome correlations}
\label{Sec:DI}

\begin{figure*}[t]
\centering
\resizebox{0.75\textwidth}{!}{\input{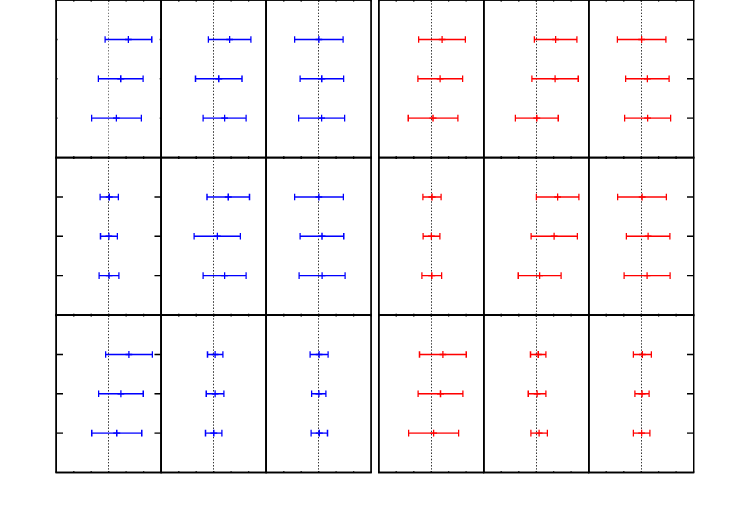}}
\caption{Differences in marginal probability distributions $\Delta P(a|x, y_{1,2})$ for Alice (blue plots on the left) and $\Delta P(b\vert x_{1,2}, y)$ for Bob (red plots on the right), evaluated for $\Pexp$ obtained from the measurement of the Bell inequality $\ImaxN{14}$ using qubits, with $\gamma = 1$ ($1\sigma$ uncertainties). The number pair $y_{1,2},x_{1,2} \in \{0,1,2\}$ refers to the pair of settings for which the difference according to \eqref{eq:NSinExpData_ax} and \eqref{eq:NSinExpData_by}, respectively, is evaluated.}
\label{fig:NonSignalingPlot_I14}
\end{figure*}

One of the most important features of  device-independent quantum information is that all conclusions --- such as the security of the distributed key, generated randomness or characteristics of the underlying state ---  are drawn directly from the observed correlations between the measurement outcomes, without invoking  any assumption about the internal workings of the device, nor the quantum state that gives rise to these correlations. Implicit in the analysis of device-independent quantum information is thus the assumption that the observed correlation is quantum realizable, cf.~\eqref{Eq:QuantumProb}, and hence non-signaling, cf.~\eqref{Eq:NS}. Moreover, for device-independent analysis to make sense, it is paramount to ensure that the experiment is free from detection loophole (see, e.g., Sev VII B1a of Ref.~\cite{Brunner:RMP}). Since our experiments clearly did not address any of the loopholes of a Bell test, the data collected, {\em strictly speaking}, cannot be used for a device-independent analysis. Rather, the analysis discussed hereafter merely serves a means for us to analyze our experimental systems -- under the assumption that \eqref{Eq:QuantumProb} holds ---  without having to trust the measurement nor preparation device (see Ref.~\cite{Rosset:PRA:2012} for a discussion of why this is relevant). In other words, our analysis in this section, though being in the spirit of device-independence, does require the additional assumptions of (\emph{a}) i.i.d., (\emph{b}) fair sampling and (\emph{c}) that any effect that could have arisen from signaling (due to the close proximity of our systems) is negligible. 

Nonetheless, due to statistical fluctuations in finite sample size, the raw correlation $\Pexp$ estimated from \eqref{Eq:RawCorrelation} essentially always deviates from the non-signaling conditions given in \eqref{Eq:NS}. In this section, we investigate the extent to which our data is compatible with the non-signaling conditions, and hence assumption (\emph{c}) stated above, and also discuss how the (signaling) raw correlations may be post-processed so that we can make full use of the measurement statistics to characterize our experimental systems.

\subsection{Signaling in the Raw Correlation}
\label{Sec:Signaling}

To gain insight into the signaling nature of our experimentally-determined correlations, we show in figure~\ref{fig:NonSignalingPlot_I12} the differences
\begin{equation}\label{eq:NSinExpData_ax}
	\Delta P(a|x,y_{1,2}):=\vert P(a\vert x, y_1) - P(a \vert x, y_2) \vert \quad y_1 \neq y_2
\end{equation}
for all $a, x, y_1, y_2 \in \{0,1,2\}$ as well as 
\begin{equation}\label{eq:NSinExpData_by}
	\Delta P(b|x_{1,2},y):=	\vert P(b\vert x_1, y) - P(b \vert x_2, y) \vert \quad x_1 \neq x_2
\end{equation}
for all $b, x_1, x_2, y \in \{0,1,2\}$ derived from $\Pexp$ for the measurement of $\S_{12}$ when $\gamma=1$.
In particular, each subplot on the left (blue) shows \eqref{eq:NSinExpData_ax} whereas each subplot on the right (red) shows \eqref{eq:NSinExpData_by}, thereby indicating the extent of signaling, respectively, from Bob to Alice and from Alice to Bob. If the non-signaling condition is respected but only deviates from \eqref{Eq:NS} due to statistical fluctuations, one expects all these differences to vanish within the statistical uncertainty. This is indeed the case for the correlation $\Pexp$ obtained during the measurement of $\S_{12}$ for $\gamma=1$ (figure~\ref{fig:NonSignalingPlot_I12}) as well as other values of $\gamma\in[0,1]$ (not shown). Likewise, the signaling nature of correlations $\Pexp$ obtained during the measurement of $\S_{14}$ for  $\gamma=1$ (figure~\ref{fig:NonSignalingPlot_I14}) as well as other values of $\gamma\in[0,1]$ (not shown) can essentially be understood via statistical fluctuations.

\begin{figure*}[t]
\centering
\resizebox{0.75\textwidth}{!}{\input{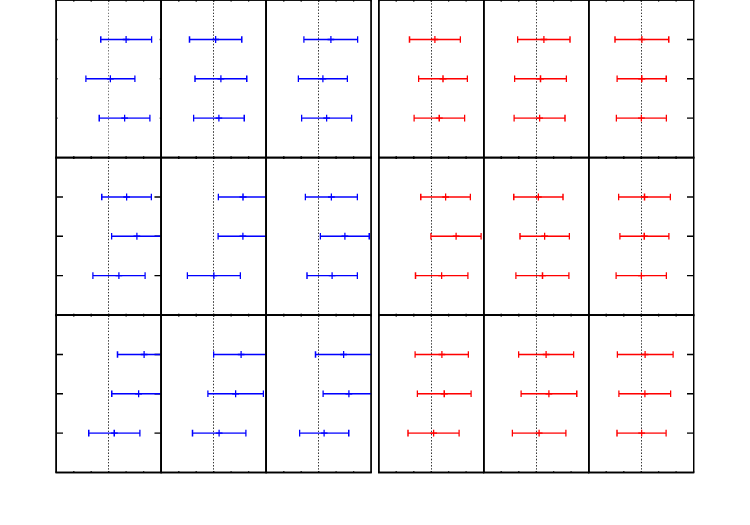}}
\caption{Differences in marginal probability distributions $\Delta P(a|x, y_{1,2})$ for Alice (blue plots on the left) and $\Delta P(b\vert x_{1,2}, y)$ for Bob (red plots on the right), evaluated for $\Pexp$ obtained from the measurement of the Bell inequality $I_3^+$ using qutrits, with $\gamma = 1$ ($2\sigma$ uncertainties). The number pair $y_{1,2},x_{1,2} \in \{0,1,2\}$ refers to the pair of settings for which the difference according to \eqref{eq:NSinExpData_ax} and \eqref{eq:NSinExpData_by}, respectively, is evaluated.}
\label{fig:NonSignalingPlot_Qutrits}
\end{figure*}

On the contrary, as can be seen in figure~\ref{fig:NonSignalingPlot_Qutrits}, even by considering a statistical uncertainty of two standard deviations, some of the differences in the conditional marginal distributions obtained for $\gamma=1$ during the measurement of $\S_0$ (i.e., for inequality $I^+_3$) do not vanish. As such, it is unlikely that the signaling observed in these correlations can be accounted for entirely through statistical fluctuations. Indeed, as separate numerical simulations indicate, the amount of signaling in these data collected for the qutrit measurement is more a consequence of the imaging arrangement. Due to the existing PSF, mentioned in section \ref{subsec:ExpSetup}., the idler and signal photon are not completely spatially separated at the SLM plane. Consequently, the modulation of one photon could influence the other photon, yielding signaling data. Moreover, due to the limited transverse spread of the total spectrum, a stronger overlap of different spectral components naturally occurs since the spacing $\Delta\omega$ of the frequency-bin pattern, given in \eqref{eq:freqBins}, is smaller for qutrits than for qubits.

\subsection{Removing Signaling by Quantum Approximation}
\label{Sec:DI-Approx}

Having established confidence that our measured correlations for $\ImaxN{12}$ and $\ImaxN{14}$ do not contradict the non-signaling conditions, we can now proceed with further analysis based on these measured correlations. To this end, if we now make the assumption that the underlying true quantum distribution also violates the measured Bell inequality by the same amount, then some simple statements, such as the amount of entanglement present in the quantum states prepared, can already be made based on the extent to which the observed correlation violates the respective Bell inequalities~\cite{Moroder:PRL:PPT}. Not surprisingly, better estimates can often be obtained if, instead, full measurement statistics are employed in the analysis (see, for instance, Refs.~\cite{NJP:Randomness1,NJP:Randomness2} in the context of randomness evaluation).

The tools that have been developed for such purposes --- such as the linear program given in ~\eqref{Eq:LP:Visibility}, or the semidefinite programs described in Refs.~\cite{DIEW,Moroder:PRL:PPT, NJP:Randomness1,NJP:Randomness2, DIWED} --- however, require explicitly that the correlation to be analyzed satisfies the non-signaling condition, \eqref{Eq:NS}. As described above, although the correlations that we obtained are compatible --- within statistical uncertainty --- with the non-signaling conditions, the raw correlations themselves do not. In order to take advantage of the full probability distribution for subsequent analysis of the source, some form of post-processing of the raw correlation would be necessary. In Ref.~\cite{NJP:Randomness2}, the authors removed the signaling in the raw correlation by projecting it into the non-signaling subspace whereas in Ref.~\cite{bernhard2014}, the authors achieved this by computing the  non-signaling distribution that is closest to the raw distribution according to the relative entropy. While the former approach works for the correlations analyzed in Ref.~\cite{NJP:Randomness2}, both these approaches suffer from the immediate risk that the non-signaling correlation derived may not be quantum realizable, cf \eqref{Eq:QuantumProb}. To minimize this risk, we will compute instead the quantum correlation nearest to the experimentally-determined correlation.

In practice, however, there are two other technical difficulties that need to be overcome before we can apply existing techniques for further analysis. Firstly, the exact characterization of the set of quantum correlations for any given Bell scenario is not known and appears to be a formidable task (see, e.g., Refs.~\cite{hierarchy1, hierarchy2}). To overcome this, we shall consider instead a converging hierarchy of relaxations to the set of quantum correlation $\Q$, such as that discussed in Refs.~\cite{hierarchy1, hierarchy2} or Ref.~\cite{Moroder:PRL:PPT} --- each of these relaxations defines a superset to $\Q$ and in the asymptotic limit, one recovers $\Q$. 

Secondly, to compute the ``nearest" quantum\footnote{Here, quantum refers to the  relaxation of $\Q$ as explained in the previous paragraph.} approximation to the experimentally-determined correlation, we need to choose an appropriate metric, i.e., a distance measure and there is apparently no unique choice to this. To facilitate computation, for the present purpose, we shall consider a distance measure based on the $\ell_1$-norm (which corresponds to minimizing the least absolute deviation). As shown in Appendix~\ref{App:SDP}, the problem of determining  a correlation $\vecP_{\Q_\ell}\in\Q_\ell$ which minimizes the distance $\vert\vert\vecP_{\Q_\ell}-\Pexp\vert\vert_{1}$ --- with $\Q_\ell$ being the $\ell$-level of the hierarchy defined in Ref.~\cite{hierarchy1} or Ref.~\cite{Moroder:PRL:PPT} --- can then be cast as a semidefinite program, which can be efficiently solved on a computer. In addition, in order to remove possible degeneracies involved in the computation, we also impose  the condition that the  distribution sought for, $\vecP_{\Q_\ell}$, reproduces the quantum value of the inequalities that we measured in the experiment, i.e., $\vec{\beta}\cdot\vecP_{\Q_\ell}=\vec{\beta}\cdot\Pexp$.

\begin{figure*}[t]
\begin{minipage}{0.49\textwidth}
\centerline{\scalebox{0.47}{\includegraphics{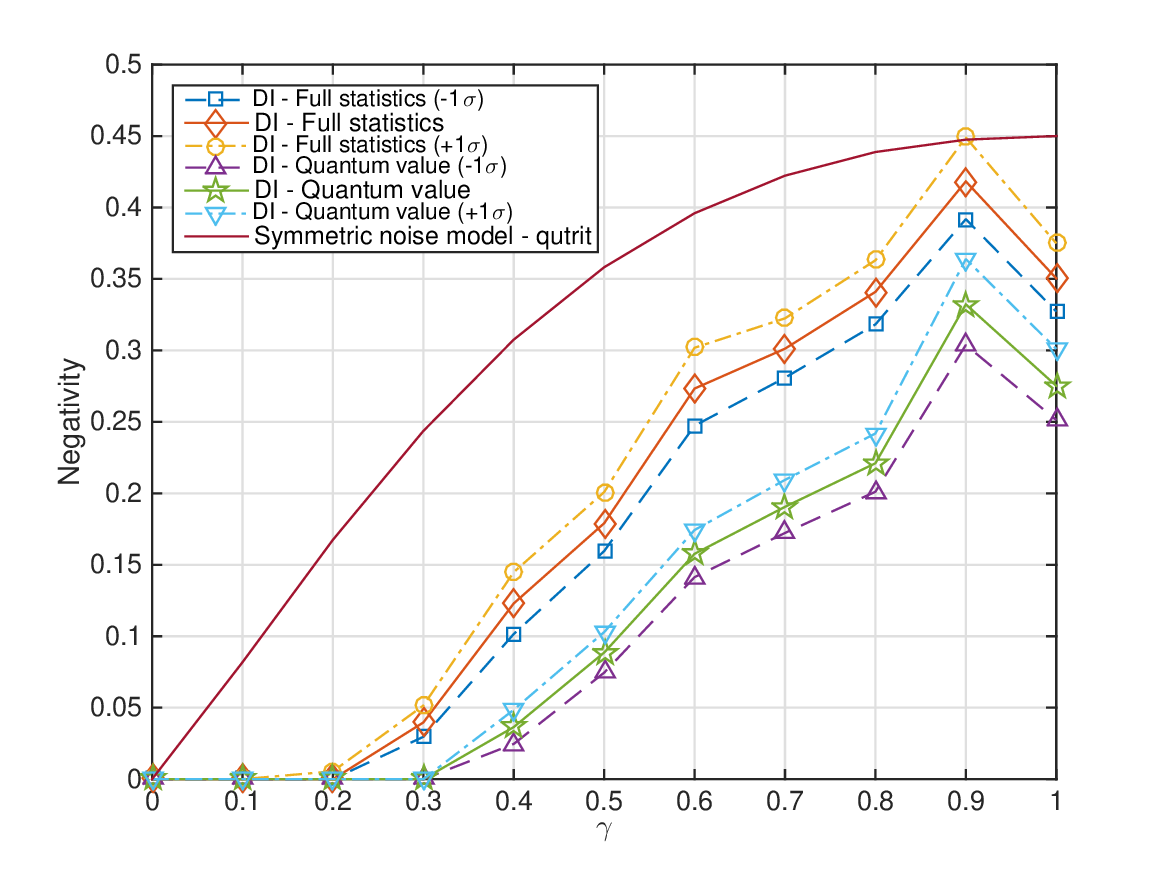}}}
\end{minipage}
\begin{minipage}{0.49\textwidth}
\centerline{\scalebox{0.47}{\includegraphics{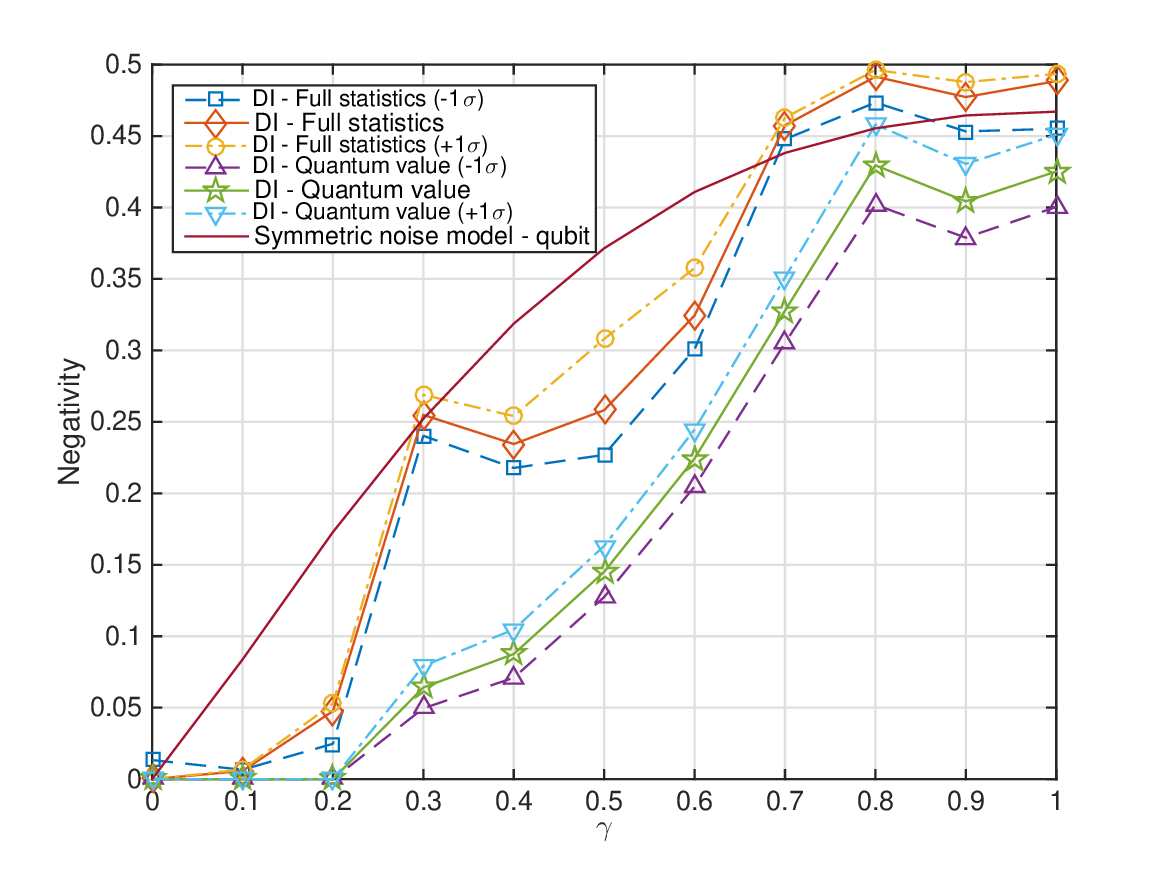}}}
\end{minipage}
\caption{\label{Fig:NegBound} 
Plots of negativity estimated directly from correlations obtained in the Bell test of inequality $\ImaxN{12}$ (left) and $\ImaxN{14}$ (right) as a function of $\gamma$ [see \eqref{eq:TwoQutritStateWithGamma} with $\gamma'=0$]. In each subplot, ``DI - Quantum value"  and ``DI - Full statistics" refer to the device-independent-inspired estimations of negativity obtained using, respectively,  {\em only} the (measured) quantum value of the corresponding Bell inequality and the full data (obtained by determining the nearest quantum approximation to the raw data). For each of these estimations, solid line assumes exactly the quantum value measured, dashed-dotted line assumes that the quantum value equals to the measured value plus  1 $\sigma$ whereas the dashed line assumes that the quantum value equals to the measured valued less 1 $\sigma$.
Also plotted are the negativities assuming a symmetric noise model [i.e., the underlying state is an isotropic state, cf. \eqref{Eq:StateVisibility}] with visibility given in the caption of figure~\ref{fig:resultsI12I14QubitAndI3Qutrit}.}
\end{figure*}

As an illustration of the above procedures, we show in figure~\ref{Fig:NegBound} an estimation of the entanglement --- measured according to negativity~\cite{Vidal:Negativity} --- present in our source computed using the technique introduced in Ref.~\cite{Moroder:PRL:PPT}.\footnote{Here, all computations are carried out assuming an intermediate-level relaxation between local level 1 and local level 2~\cite{Moroder:PRL:PPT}, using a $\chi$ matrix of the size of $289\times289$.}   Evidently, the negativity lower bound obtained from the (approximated) full data are generally higher than that arising from the quantum violation of the respective Bell inequalities. In fact, as can be seen in figure~\ref{Fig:NegBound}, the negativity estimated from the full measurement data are sometimes even higher than that obtained from the {\em device-dependent} estimation assuming a symmetric noise model. Note also that for each of these  Bell experiments, the measured correlations for certain small values of $\gamma$ are not good enough to violate  the corresponding Bell inequalities. However, when the full data is taken into account, we are sometimes able to obtain non-trivial estimates of the entanglement present in the underlying state.\footnote{Whenever this happens, the corresponding correlation $\vecP$ has to be outside $\L$, albeit this is not reflected by the quantum value of the measured Bell inequality. }

To test the reliability of the proposed method, we have also computed the negativity (lower bound) by assuming that the true quantum value is the measured quantum value plus (minus) one standard deviation [see dotted (dashed) line in figure~\ref{Fig:NegBound}]. As can be seen in the figure, except for one instance, namely for $\ImaxN{14}$ and when $\gamma=0$, all negativities estimated using the actual value fit well between those estimated assuming the shifted values. For this exceptional instance, the negativity estimated from the full data (and the assumption that the true quantum value is the measured value less one $\sigma$) turns out to give an even larger (and {\em nonzero}) value compared with the case when the assumed quantum value is the measured value (or with one standard deviation above). A possible cause for this somewhat unexpected result is that even with an assumption of the true quantum value, there may be more than one $\vecP_{\Q_\ell}\in\Q_\ell$ that minimizes the $\ell_1$ distance to the raw correlation, and the solver happens to have returned a  $\vecP_{\Q_\ell}$ that gives the highest value of negativity (when the assumed quantum value is the smallest of the three cases considered). Another unusual feature associated with this exceptional instance is that, by right, for $\gamma=0$, our system should have produced a separable state and is thus not capable of violating any Bell inequality (nor producing correlation to be associated with non-trivial negativity). Thus, if the estimation that we have obtained using the (approximated) full data (and the assumption that the true quantum value is indeed the measured value minus one $\sigma$) is legitimate, then this conclusion obtained from the measured data nicely illustrates how our trust in the measurement and preparation devices may be misguided.

\section{Discussion}
\label{Sec:Conclusion}

In this work, starting from the Bell inequality $I^+_3$ presented in Ref.~\cite{Liang:PRA:2009} and analyzing its quantum violation for a family of two-qutrit states, we obtain several novel facet-defining Bell inequalities for the Bell scenario $\{[3\,3\, 3]\, [3\, 3\, 3]\}$. Interestingly, all these newly obtained Bell inequalities that are {\em not} reducible to simpler Bell scenarios can already be violated maximally using entangled two-qubit states, including a Bell state. In contrast, some of these Bell inequalities that are reducible to Bell inequalities involving fewer number of outcomes require entangled two-qutrit states to achieve maximal quantum violation. These results show, once again (see e.g., Ref.~\cite{Pal:042105,Pal:PRA:022120,Grandjean:PRA:2012}),  that in analyzing the quantum violation of given Bell inequalities, the widely-employed wisdom of setting the local Hilbert space dimension to be the same as the number of measurement outcomes is unfounded.

The quantum violations for two of these newly obtained Bell inequalities, together with $I^+_3$, are experimentally demonstrated using, respectively, a one-parameter family of entangled two-qubit and two-qutrit states. These are Bell inequalities that are maximally violated, respectively, by a maximally entangled two-qutrit state, a maximally entangled two-qubit state and a partially entangled two-qubit state in conjunction with genuine POVM. Not surprisingly, the raw measurement statistics deviate from ideal quantum correlations which ought to satisfy the non-signaling condition. In the case of qubit measurements, these deviations are compatible with what one would expect from finite-size effects (quantifiable through statistical uncertainty) whereas in the qutrit case, unfortunately, such deviations cannot be entirely accounted for using finite statistics. Separate numerical simulations nevertheless show that this last observation probably stems from an imperfect implementation of the qutrit measurements, due to limited optical resolution at the SLM plane. As a result, in addition to giving a much lower signal-to-noise ratio, unwanted interferences between specific bins are more likely to occur, yielding the appearance of signaling in the data sets.  

Naturally, given that the raw correlation exhibits signaling, one may raise doubts on any conclusions drawn directly from these correlations. In the event when the deviation (from non-signaling) is accountable for by statistical fluctuations, such doubts may be alleviated, thus allowing one to proceed with a device-independent-inspired analysis under the assumptions that (1) the source and the measurements are i.i.d, (2) the counts registered represent a fair sample of all the measurement events. For convenience, we also make the commonly-adopted assumption that the true underlying quantum distribution gives rise to the same measured value of Bell inequality violation. Building on these assumptions, as we discussed in section~\ref{Sec:DI}, stronger statement may be made if one looks for the nearest quantum approximation to the experimentally-determined correlation, and continue the analysis from there. The essence of this approach is thus in the same spirit as the familiar approach of quantum state tomography using maximum likelihood state estimation --- the main difference being that the ``tomography" is now applied to the raw correlation of measurement outcomes obtained in experiments.

Let us now comment on some possibilities for future research. Experimentally, ongoing work aims at achieving a better optical resolution, a higher signal-to-noise ratio setup and an experimental arrangement in which the idler and signal photon can be manipulated individually. The finite optical resolution at the SLM plane leads to the aforementioned, unwanted overlap between spectral components and therefore forbids us from having well-separated subsystems. This overlap can be reduced by incorporating a prism based pulse shaper in which the single lens imaging system is replaced by a 4f-imaging arrangement, the latter providing a significantly better imaging resolution quality. Further, since SFG in a single nonlinear crystal is a local detection process with both photons taking the same optical path through the experimental setup, another goal in future is to have a detection scheme based on two delocalized nonlinear crystals. The up-conversion process between the idler and signal photon is then triggered by a strong seed pulse and the up-converted photons are finally detected in coincidence electronically~\cite{kuzucu2008}. This scheme enhances the signal-to-noise ratio compared to the detection process presented in this work. Moreover, it would allow for manipulating the idler and signal photon in separated paths by means of a second SLM. 

Coming back to the theoretical side, one of the original goals of the present work is to look for a robust (i.e., with good noise tolerance) Bell inequality that can be used for the self-testing of a maximally entangled two-qutrit state. While $\I^+_3$ from Ref.~\cite{Liang:PRA:2009} may serve, in principle, as such a candidate, its rather poor visibility does not make it a very attractive candidate for practical purposes. On the other hand, although  we did obtain a number of previously unknown facet-defining Bell inequalities, none of them turn out to be suitable for this purpose either. The search for such a candidate thus remains as an opened problem. 

Secondly, let us remark that the signaling nature of raw correlations observed in a Bell test is an issue that has been largely overlooked in the studies of device-independent quantum information. In this work, we have discussed a method alternative to that of Refs.~\cite{bernhard2014,NJP:Randomness2} in order to deal with this generic problem faced in the analysis of real experimental data. While the method has worked more or less as expected, we have not thoroughly investigated its reliability. For instance, it seems conceivable that  in the generic situation, there may be more than one correlation $\vecP_{\Q_\ell}\in\Q_\ell$ that gives the same distance to the raw correlation $\Pexp$ even after we impose the condition that $\vec{\beta}\cdot\vecP_{\Q_\ell}=\vec{\beta}\cdot\Pexp$. How would such non-uniqueness affect any  device-independent conclusions, or can we remove this problem via other choice(s) of distance measures or imposing other natural constraints? In addition, given that $\Q_\ell$ only approaches $\Q$ as $\ell\to\infty$, how would this difference from $\Q$ for any finite value of $\ell$ affect the validity of our estimation? And even if we can overcome these difficulties, how do we properly translate the statistical uncertainty obtained in the raw correlation into the quantities of interest, such as the entanglement of the underlying state, or the randomness extractable from a certain outcome? Although the n\"aive approach that we adopted towards the end of section~\ref{Sec:DI} does seem to suggest a confidence region of the estimated negativity for most of the cases analyzed, it is clear from our observations that a more thorough investigation needs to be carried out. Note that similar questions have also been discussed in the context of quantum state tomography where some solutions have been proposed, see, for instance Ref.~\cite{FiniteTomography}. Can these tools be adapted for device-independent analysis? These questions are clearly of paramount importance for the future implementation of device-independent quantum information protocols, thus leaving us plenty to ponder upon for future research.

\begin{acknowledgments}
YCL would like to thank the Group of Applied Physics at the University of Geneva for allowing us to access their computational resource in order to bound the maximal quantum violation of $\ImaxN{19}$, and to Jean-Daniel Bancal for sharing his Matlab programs in order to determine if a Bell inequality can be cast in a form involving only correlators. This research was supported by the  Swiss National Science Foundation (NCCR-QSIT and grant number PP00P2$\_$133596), the Hasler Foundation, through the project ``Information-Theoretic Analysis of Experimental Qudit Correlations``, the Ministry of Education, Taiwan, R.O.C., through ``Aiming for the Top University Project" granted to the National Cheng Kung University (NCKU), and the Ministry of Science and Technology, Taiwan (grant number 104-2112-M-006-021-MY3).
\end{acknowledgments}

\appendix

\section{Linear program}
\label{App:LP}

For any given correlation $\vec{P}_G=\{P_G(a,b|x,y)\}$ of measurement outcomes, one can determine its white-noise visibility $\vcr$ with respect to $\L$ via the following linear program
\begin{subequations}\label{Eq:LP:Visibility}
\begin{align}
	&\max \qquad\qquad\qquad v\\
	&\text{s.t. }\quad v\, \vec{P}_G + (1-v)\vec{P}_\id=\sum_{i=1}^n p_i \vec{P}^{\L}_{\mbox{\tiny Ext},i},\\
	&\qquad\quad v\ge 0,\quad p_i\ge 0\,\,\,\forall\,\,\, i,\quad  \sum_i p_i=1,
\end{align}
\end{subequations}
where $\vec{P}_\id$ is the white noise correlation, $\vec{P}^{\L}_{\mbox{\tiny Ext},i}$ is the $i$-th extreme point of $\L$ ---each extreme point only has entries $0,1$--- $p_i$ is the weight associated with each of these extreme points and $n$ is the total number of extreme points of $\L$.

Now, recall from~\cite{Book:CVX} that every linear program of the form
\begin{equation}
\begin{split}
	&\max \qquad {\bf c^T}{\bf x}\\
	&\text{s.t. }\qquad A{\bf x}={\bf b},\\
	&\qquad\qquad {\bf x}\succeq 0,
\end{split}
\end{equation}
has a dual program of the form
\begin{equation}
\begin{split}
	&\min \qquad {\bf b^T}{\bf y}\\
	&\text{s.t. }\qquad A^T{\bf y}\succeq{\bf c},
\end{split}
\end{equation}
where ${\bf x}$ and ${\bf y}$ are respectively column vectors collecting all optimization variables of the primal and the dual linear program, while $\succeq$ denotes entry-wise inequality.

The dual program corresponding to \eqref{Eq:LP:Visibility} is thus
\begin{subequations}\label{Eq:LP:Dual}
\begin{align}
	&\min \qquad \vec{P}_\id\cdot\vec{y}+y_0\\
	&\text{s.t. }\quad \vec{P}^{\L}_{\mbox{\tiny Ext},i}\cdot\vec{y}+y_0\ge 0\quad\forall\quad i\in\{1,\ldots,n\},\label{Eq:LP:BI-1}\\
	&\quad\quad\,\,\,\,\, \vec{P}_G\cdot\vec{y} \le \vec{P}_\id\cdot\vec{y}-1\label{Eq:LP:BI-2}.
\end{align}
\end{subequations}
Note that the linear inequality suggested by the dual linear program 
\begin{equation}\label{Eq:LP-BI}
	\vec{P}\cdot\vec{y}+y_0\ge 0,
\end{equation}
is a Bell inequality, since \eqref{Eq:LP:BI-1} guarantees that it holds for all extreme points of $\L$, and hence for all  $\vec{P}$ resulting from convex combinations thereof.

When strong duality holds, the linear program return an optimum value satisfying $v=\vcr=\vec{P}_\id\cdot\vec{y}+y_0=\vec{P}_\id\cdot\vec{y}^*+y_0^*$, where $(\vec{y}^*,y_0^*)$ are the optimal dual variables. In particular, if the optimum value $\vcr$ is less than 1, we see from \eqref{Eq:LP:BI-2} that the Bell inequality given by \eqref{Eq:LP-BI} is violated. Hence, by solving the linear program given in \eqref{Eq:LP:Visibility}, we not only obtain the critical white-noise visibility $\vcr$, but also (in the case when $\vcr*<1$) a Bell inequality, cf. \eqref{Eq:LP-BI}, certifying that the given correlation is indeed outside $\L$.

\section{Optimal Measurement for $\ImaxN{12}$}
\label{App:OptMeas12}

Here, we provide a set of measurements that can be used in conjunction with the two-qubit state
\begin{equation}
	\ket{\psi_{\mbox{\tiny max}}^{12}}=0.7258\ket{00}+0.6879\ket{11}
\end{equation}
in order to achieve the maximal quantum violation of $\ImaxN{12}$. Specifically, let us consider projective POVM elements such that $M_{a|x}=\proj{\phi^A_{a|x}}$ and $M_{b|y}=\proj{\phi^B_{b|y}}$, where $\ket{\phi^A_{a|x}}$ and $\ket{\phi^B_{b|y}}$ are column vectors which ---when expressed in the $\{\ket{i}\}_{i=0}^2$ basis --- have entries given, respectively, by the $a$-th and the $b$-th column vector of the matrix $\Omega^{\mbox{\tiny A}}_{x}$ and $\Omega^{\mbox{\tiny B}}_{y}$ specified below:
\begin{align*}
\Omega^{\mbox{\tiny A}}_{1}&=\left(\begin{array}{ccc} 0.9835 & 0 & -0.1809\\ 0.1809 & 0 & 0.9835\\0 & 1 & 0\end{array}\right),\\
\Omega^{\mbox{\tiny A}}_{2}&=\left(\begin{array}{ccc} 0.8826 & -0.4642 & 0.0742\\ 0.4626 & 0.8857 & 0.0389\\ 0.0492-0.0678i & 0 & -0.5851+0.8066i\end{array}\right),\\
\Omega^{\mbox{\tiny A}}_{3}&=\left(\begin{array}{ccc} -0.9117 & 0.4108 & 0\\ 0.4108 & 0.9117 & 0\\ 0 & 0 & 1\end{array}\right),\\
\Omega^{\mbox{\tiny B}}_{1}&=\left(\begin{array}{ccc} 0.9974 & 0 & -0.0721\\ 0.0721 & 0 & 0.9974\\0 & 1 & 0\end{array}\right),\\
\Omega^{\mbox{\tiny B}}_{2}&=\left(\begin{array}{ccc} -0.8799 & 0.2668 & -0.3932\\ 0.2436 & 0.9637 & 0.1089\\ -0.4080-0.0004i & 0 & 0.9130+0.0008i\end{array}\right),\\
\Omega^{\mbox{\tiny B}}_{3}&=\left(\begin{array}{ccc} 0.7478 & -0.6639 & 0\\ 0.6639 & 0.7478 & 0\\ 0 & 0 & 1\end{array}\right).
\end{align*}

To obtain a set of genuine POVM which reproduces this maximal quantum violation, it suffices to perform the projection $\proj{0}+\proj{1}$ onto each of these $M_{a|x}$ and $M_{b|y}$ defined in the qutrit space to obtain a set of $M_{a|x}$ and $M_{b|y}$ defined in the qubit space.

\section{Semidefinite program}
\label{App:SDP}

Recall from~\cite{Moroder:PRL:PPT} that at any given level (say, $\ell$) of the hierarchy considered therein, we consider a matrix $\chi_\ell[\rho]$ that can be decomposed as
\begin{equation}\label{Eq:Decomposition}
	\chi_\ell[\rho]=\sum_{\vec{a},\vec{x}} P(\vec{a}|\vec{x})\,F^\ell_{\vec{a},\vec{x}} + \sum_v u_v F^\ell_v,
\end{equation}
i.e., into one fixed part that linearly depends on the measurable quantities $\{P(\vec{a}|\vec{x})\}$, and a complementary (orthogonal) part that is known only if the underlying state $\rho$ and the measurement giving rise to the correlation $\{P(\vec{a}|\vec{x})\}$ is known; in ~\eqref{Eq:Decomposition}, $F^\ell_{\vec{a},\vec{x}}$ and $F^\ell_v$ are some fixed, symmetric, Boolean matrices~\cite{Moroder:PRL:PPT}.

For any observed raw correlation $\Pexp$, a correlation $\vecP\in\Q_\ell$ which minimizes the least absolution deviation $||\vecP(\vec{a}|\vec{x})-\Pexp(\vec{a}|\vec{x})||_{1}$ (i.e., according to the $\ell_1$ norm) can be obtained by solving the following semidefinite program
\begin{align}
	\min_{\{\vecP(\vec{a}|\vec{x})\},\{u_v\}}\quad 
	&\sum_{\vec{a},\vec{x}}D(\vec{a}|\vec{x}),\label{Eq:SDPMaxViolation}\\
	{\rm s.t.} \quad\qquad &\chi^{\phantom{T_j}}_\ell\!\!\![\rho]\ge 0, \nonumber\\
		 				    &\vecP(\vec{a}|\vec{x})-\Pexp(\vec{a}|\vec{x})\le D(\vec{a}|\vec{x})
						    \quad\forall\vec{a},\vec{x}, \nonumber\\
		 				    &\Pexp(\vec{a}|\vec{x})-\vecP(\vec{a}|\vec{x})\le D(\vec{a}|\vec{x})
						    \quad\forall\vec{a},\vec{x}, \nonumber
\end{align}
where $D(\vec{a}|\vec{x})$ is an array of auxiliary variables having the same dimension as $\vecP$. Notice that even if we include in \eqref{Eq:SDPMaxViolation} the additional constraint that $\vecP(\vec{a}|\vec{x})$ must reproduce the quantum value of some Bell expression specified by $\vec{\beta}$, i.e., 
\begin{equation}
	\vec{\beta}\cdot\vecP=\vec{\beta}\cdot\Pexp,
\end{equation}					    
the resulting optimization problem remains as a semidefinite program, as this constraint is  linear in the optimization variables $\vecP(\vec{a}|\vec{x})$.

\end{document}


\title{Corrigendum to ``Bipartite Bell inequalities with three ternary-outcome measurements
--- from theory to experiments" [New J Phys. {\bf 18}, 035001 (2016)]}

\author{Sacha Schwarz}
\affiliation{Institute of Applied Physics, University of Bern, 3012 Bern, Switzerland.}
\author{B\"anz Bessire}
\affiliation{Institute of Applied Physics, University of Bern, 3012 Bern, Switzerland.}
\author{Andr\'e Stefanov}
\affiliation{Institute of Applied Physics, University of Bern, 3012 Bern, Switzerland.}
\author{Yeong-Cherng~Liang}
\email{ycliang@mail.ncku.edu.tw}
\affiliation{Department of Physics, National Cheng Kung University, Tainan 701, Taiwan.}

\date{\today}


\maketitle

In our article~\cite{BellExpNJP}, we claimed to have given the first example of a facet-defining Bell inequality where a genuine positive-operator-valued (POVM) measure is relevant. Specifically, we claimed that any quantum realization of the maximal quantum violation of the Bell inequality $\Imax_{12}$ in the (minimal) qubit subspace necessarily requires the implementation of a genuine, nonprojective POVM. Recently, it has been brought to our attention by Armin Tavakoli that the maximal quantum violation of this Bell inequality can also be achieved using projective measurements. 

Specifically, consider~\cite{Tavakoli} the following POVM elements $\{M^A_{a|x}\}_{a,x}$ and $\{M^B_{b|y}\}_{b,y}$, respectively, for Alice's and Bob's measurements:
\begin{equation}
\begin{split}
	M^A_{1|0}=M^A_{2|1}=M^A_{2|2}=M^B_{1|0}=M^B_{2|1}=M^B_{2|2}=\mathbf{0}_2,\\
	M^A_{0|x}=\tfrac{1}{2}\left(\id_2+\hat{\eta}^A_x\cdot\vec{\sigma}\right),\quad
	M^B_{0|y}=\tfrac{1}{2}\left(\id_2+\hat{\eta}^B_y\cdot\vec{\sigma}\right),\\
\end{split}
\end{equation}
where $\mathbf{0}_2$ and $\id_2$ are, respectively, the zero operator and the identity operator acting on a qubit Hilbert space, $\vec{\sigma}=(\sigma_x,\sigma_y,\sigma_z)$ is the vector of Pauli matrices, and
\begin{align*}
	\hat{\eta}^A_0=\left(\begin{array}{r}
	0\\ 0\\ 1
	\end{array}\right),\quad
	&\hat{\eta}^A_1=\left(\begin{array}{r}
	-0.5009\\-0.2634\\ 0.8244
	\end{array}\right),\quad
	\hat{\eta}^A_2=\left(\begin{array}{r}
	0.8259\\0.4400\\ 0.3525
	\end{array}\right),\\		
	\hat{\eta}^B_0=\left(\begin{array}{r}
	0.1916\\ 0.1328\\ 0.9724
	\end{array}\right),\quad
	&\hat{\eta}^B_1=\left(\begin{array}{r}
	0.6907\\ -0.3933\\ 0.6069
	\end{array}\right),\quad
	\hat{\eta}^B_2=\left(\begin{array}{r}
	-0.7767\\ 0.4117\\ 0.4767
	\end{array}\right).		
\end{align*}
It is straightforward to verify that together with the quantum state 
$$\ket{\psi}=(0.7244+0.0083 \ii)\ket{00}+(0.0066-0.0131 \ii)\ket{01}+(0.0054-0.0062\ii)\ket{10}+0.6891\ket{11},$$ 
one obtains 2.5820, which reproduces (within the numerical precision of our computation) the maximal quantum violation of $\Imax_{12}$. Our claim is thus flawed. Our mistake arose from a flaw in the numerical computation of the maximal possible quantum violation when some of the POVM elements are assumed to be the zero operator.

Consequently, it remains an opened problem whether there exists a facet-defining Bell inequality whose maximal quantum violation can only attained by employing a nonprojecitve measurement when one restricts to the smallest Hilbert space where this violation is achievable.